# Deterministic Localization of Strain-induced Single-photon Emitters in Multilayer GaSe

Weijun Luo,[1] Alexander Puretzky,[2*] Benjamin Lawrie,[2,3*] Qishuo Tan,[1] Hongze Gao,[1] Zhuofa Chen,[5] Alexander Sergienko,[5,6] Anna Swan,[5,6] Liangbo Liang,[2] Xi Ling[1,4,6*]

[1] Department of Chemistry, Boston University, Boston, MA 02215, United States
[2] Center for Nanophase Materials Sciences, Oak Ridge National Laboratory, Oak Ridge, TN 37831, United States
[3] Materials Science and Technology Division, Oak Ridge National Laboratory, Oak Ridge, TN 37831, United States
[4] Division of Materials Science and Engineering, Boston University, Boston, MA 02215, United States
[5] Department of Electrical Engineering, Boston University, Boston, MA 02215, United States
[6] The Photonics Center, Boston University, Boston, MA 02215, United States

*To whom the correspondence should be addressed. Email address: xiling@bu.edu; puretzkya@ornl.gov; lawriebj@ornl.gov

**Abstract**
Nanoscale strain has emerged as a powerful tool for controlling single-photon emitters (SPEs) in atomically thin transition metal dichalcogenides (TMDCs)(*1*, *2*). However, quantum emitters in monolayer TMDCs are typically unstable in ambient conditions. Multilayer two-dimensional (2D) TMDCs could be a solution, but they suffer from low quantum efficiency, resulting in low brightness of the SPEs. Here, we report the deterministic spatial localization of strain-induced single-photon emitters in multilayer GaSe by nanopillar arrays. The strain-controlled quantum confinement effect introduces well-isolated sub-bandgap photoluminescence and corresponding suppression of the broad band edge photoluminescence. Clear photon-antibunching behavior is observed from the quantum dot-like GaSe sub-bandgap exciton emission at 3.5 Kelvin. The strain-dependent confinement potential and the brightness are found to be strongly correlated, suggesting a promising route for tuning and controlling SPEs. The comprehensive investigations of strain-engineered GaSe SPEs provide a solid foundation for the development of 2D devices for quantum photonic technologies.

**Introduction**

Over the past two decades, the scalable and controllable generation of SPEs with high brightness, purity, and indistinguishability has emerged as a critical requirement for developing photonic quantum technologies(*3*, *4*). Recently, SPEs in 2D layered materials have drawn increasing interest because of the potential for deterministic generation and integrability with photonic devices(*5*). SPEs arising from defect-bound excitons have been observed in monolayer TMDCs such as $WSe_2$(*6–8*) with nanoscale strain providing deterministic control of SPEs(*1*, *9*). Nevertheless, SPEs in monolayer TMDCs are sensitive to ambient instabilities that can cause a significant variation in the observed quantum optical properties(*9*, *10*). On the other hand, multilayer TMDCs, which exhibit better stability, have indirect bandgaps and do not exhibit intense photoluminescence (PL)(*11*). As a result, research targeting control of SPEs in environmentally robust multilayer TMDCs has remained limited.

In contrast to the decreasing quantum yield of TMDCs with increasing numbers of layers, the band edge PL of layered GaSe increases dramatically with increasing numbers of layers(*12*), suggesting that multilayer GaSe is a potential host for environmentally robust SPEs. Although multilayer GaSe has an indirect bandgap of 2.05 eV, there is a very small energy difference (< 0.1 eV) between its indirect and direct bandgaps(*13*, *14*), allowing for possible strain-induced indirect-to-direct bandgap transitions(*15*). Prior PL studies of GaSe have shown that strain manipulation can significantly enhance its band edge PL emission by one order of magnitude(*16*, *17*). However, reports of GaSe hosting SPEs are limited to the observations of randomly distributed SPEs that have been attributed to strain-bound excitons caused by inhomogeneous selenium clusters in the material(*18*, *19*). To date, deterministic control of SPEs in multilayer GaSe with local strain has not been demonstrated.

In this work, we report an order-of-magnitude enhancement of the sub-bandgap PL emission from strained multilayer GaSe flakes (with thicknesses between 20 and 50 nm) on $SiO_2$ nanopillar arrays. Strong PL from strain-confined excitons and biexcitons with emission energies between 1.6 eV and 2.0 eV is observed at a temperature of T = 3.5 K. Notably, the exciton PL of GaSe at most pillar locations exhibits photon antibunching characteristic of single-photon emission. By varying the incident laser power density, we show that the biexciton emission can be suppressed, resulting in improved single-photon purity. Moreover, the strain distribution around the pillar area is analyzed and used to quantify the effect of the strain-dependent quantum confinement on the emitter brightness with Pearson correlation analysis(*20*). It clearly shows the positive correlation between strain, confinement potential, and brightness of the emitters. The control of confinement potential through localized strain in these nanopillar geometries thus provides a key tool for fundamental studies of SPE photo-physics in multilayer GaSe and for the development of 2D quantum photonic devices.

**Results**

As shown in Fig. 1A, multilayer GaSe flakes (with a thickness of ~ 30 nm) are first mechanically exfoliated from high-quality ε-type GaSe bulk crystals (the crystal structure of a monolayer and Raman spectrum of GaSe bulk crystal are shown in Supplementary Fig. S1) and then transferred to a Si substrate with pre-patterned $SiO_2$ nanopillar arrays (see more details on the sample preparation and pillar topography in Supplementary Figs. S2 and. S3, respectively). For simplification, we illustrate monolayer GaSe instead of multilayer GaSe in the schematic shown in Fig. 1A. The $SiO_2$ nanopillar arrays were designed with three diameters of ~ 150 nm, ~ 200 nm, and ~ 250 nm, and an average height of ~ 100 nm (Supplementary Figs. S2 - S4). The nanopillar diameters were chosen to be less than half of the excitation wavelength of 532 nm to drive quantum confinement effects(*2*). The GaSe flake is tented uniformly on most pillars without piercing or breakage (as shown in the AFM image of Supplementary Fig. S5). Since the PL intensity of GaSe increases drastically for thickness between 20 nm and 50 nm(*12*), the thicknesses of all GaSe flakes studied in this work are within this range. Fig. 1B shows an extracted PL intensity map (measured at T = 3.5 K with a laser power density of 2 $\mu W/\mu m^2$; the complete map is shown in Supplementary Fig. S6) of a prototypical GaSe flake (thickness, ~ 30 nm) transferred onto a $SiO_2$ nanopillar array (as shown in the inset image). At this moderately high laser power density, several PL bands are observed spanning 600 – 750 nm (Supplementary Fig. S7). The ratio of the strain localized GaSe PL to the band edge emission $X_{Eg}$ is significantly enhanced by two orders of magnitude on the pillars. At a much lower laser power density of 44 $nW/\mu m^2$, PL spectra measured at T = 3.5 K on the pillars exhibited a few strong and well-separated features without the broad background or additional defect-related bands that are often observed in $WSe_2$(*7*) and hBN(*21, 22*). For example, in Fig. 1C, the PL spectrum collected from GaSe on a pillar shows a single peak $X_s$ at ~1.90 eV (~ 652 nm), about ~ 0.15 eV below the well-studied GaSe band edge emission $X_{Eg}$ at 2.05 eV(*23–26*). Note that the emission energies of GaSe SPEs measured here are consistent with previous work on GaSe SPEs associated with strain developed on GaSe film on randomly distributed selenium clusters(*18*).

To evaluate the anti-bunching performance of this SPE, Hanbury Brown-Twiss (HBT) interferometry was used to characterize the second-order correlation function $g^{(2)}(\tau)$ of the emission peak $X_s$ at ~1.90 eV (~ 652 nm). As shown in Fig. 1D, anti-bunching with $g^{(2)}(0)$ = 0.32 ± 0.02 and a decay time-constant of ~ 4.1±0.7 ns is observed (the fitting details are discussed in Section 6 in the supplementary information). We also identify GaSe SPEs on the pillars with energies of 1.65 eV – 2.0 eV and time constants of 0.7-15.5 ns as summarized in Supplementary Figs. S8 and S9. However, the observed sub-bandgap transitions exhibit varying single photon purity; as shown in Supplementary Fig. S10, some emitters with $g^{(2)}(0) > 0.5$ are also observed. The GaSe SPEs exhibited an average wandering value of ~ 0.5 nm under different incident laser power densities (Supplementary Fig. S11), which is much better than values previously reported for $WSe_2$ (~2 nm)(*27*) and hBN(*28*) (~ 5 nm) SPEs. No significant blinking or bleaching are observed for all the SPEs studied.

Note that the incident laser power density is critical in the single photon purity. Previous studies of SPEs in quantum dots(*29*), carbon nanotubes,(*30*) color centers in diamond(*31*), and $WSe_2$(*32*) have demonstrated that a low excitation power density is required to minimize the emergence of additional excited states that can reduce the SPE purity. Therefore, to probe the interplay between brightness and purity for this 1.90 eV SPE, we performed PL and HBT interferometry measurements with various incident laser power densities. Fig. 2A shows five high-resolution spectra collected with incident laser power densities of 22 nW/$\mu m^2$ to 131 nW/$\mu m^2$: a single symmetric exciton peak $X_s$ (at ~ 1.905 eV) is present at laser power densities below 87 nW/$\mu m^2$; a weaker biexciton peak XX (at ~ 1.916 eV) emerges on the higher energy side of the excitonic peak $X_s$ (energy shift $|\Delta E|$ ~ 11 meV) and becomes relatively prominent for an incident laser power density of ~ 131 nW/$\mu m^2$. Note that the biexciton formation with a binding energy of ~ 2 meV was reported for GaSe(*33*). Fig. 2B illustrates HBT interferometry measurement results as functions of incident laser power densities: $g^{(2)}(0)$ remains less than 0.5 when the incident laser power density is in the range of 60-150 nW/$\mu m^2$ (Supplementary Fig. S12A-E). With the increase of the laser power density, the $g^{(2)}(0)$ increases and the SPE purity reduces, due to the increased contribution of the biexciton. Also, as the impact of various environmental fluctuations becoming more significant at low laser power density region (<60 nW/$\mu m^2$), the $g^{(2)}(0)$ value does not get improved. (Supplementary Fig. S12I). The fitted decay time constants generally decrease with increasing laser power density, resulting in increased brightness. Hence, our results indicate that the brightness increase under a higher incident laser power density does not necessarily improve the SPE purity due to the emergence of the biexciton features. Consistent results were also observed on another SPE at ~ 1.79 eV (693 nm), and the additional power density-dependent photon-antibunching measurements are shown in Supplementary Figs. S13 and S14. Fig. 2C illustrates a PL spectrum of the same emitter at 1.90 eV (652 nm) acquired with a higher incident laser power density of 300 nW/$\mu m^2$, where there is no observed anti-bunching. Four peaks are observed and labeled as $XX_1$, $X_S$, $XX_2$ and $XX_3$ at ~ 1.904 eV, ~ 1.907 eV, ~ 1.910 eV, and ~1.916 eV, respectively. Note that previous works demonstrated that the GaSe PL peaks could be classified into excitons[15] and biexcitons(*33*) according to the sub-linear and super-linear power-law scaling of the PL intensity. Fig. 2D shows the integrated counts (C) of different peaks as a function of incident laser power densities (P) plotted in a double-logarithmic plot and fitted using the power-law function: $C \propto P^x$. The measured slope, $x$, can be used to classify GaSe excitons(*18, 34*) and biexcitons(*18, 35*) because excitons and biexcitons exhibit slopes of less than one and larger than one, respectively. The slope of the $X_S$ peak is 0.62, so $X_S$ is assigned to an exciton feature; $XX_1$, $XX_2$ and $XX_3$ have slopes greater than one, and they are identified as biexciton features.

Interestingly, some biexcitons (i.e., $XX_1$ and $XX_2$) are observed with higher energy than the exciton, implying negative binding energies. Across all the results, we observe both negative binding energies (Figs. 2C and 3A) and positive biexciton binding energies (Supplementary Figs. S10A and S10C). Negative biexciton binding energies resulting from biexciton antibonding or exciton-exciton repulsion have been primarily observed in quantum dot systems(*36–39*). Hence, it could be inferred that the nanopillar arrays cause the formation of a quantum confinement potential, resulting in quantum dot-like GaSe SPEs. The localized strain in GaSe at different pillars introduces different confinement potential(*40*), charge-carrier localization(*41*), and charge separation(*42*), leading to the different observed exciton and biexciton binding energies as well as different required laser power densities for the formation of excitons.

Polarization-dependent measurements on an SPE and the corresponding biexciton are performed to better understand the relationship between a exciton and biexciton. Fig. 3A shows the PL spectra of an emitter $X_s$ (at 1.99 eV, ~ 620 nm) acquired with a horizontally polarized excitation laser in parallel- and cross-polarized collection configurations. The 2.02 eV (~ 615 nm) peak was recognized as a biexciton feature from the power density-dependence measurements (Supplementary Fig. S15B). In Fig. 3B, least squares fitting of the second-order correlation function produces a fitted value of $g^{(2)}(0) = 0.46 \pm 0.03$. Fig.

3C illustrates the integrated polarization-dependent PL from both the exciton $X_s$ and the biexciton XX. They share the same polarization response, suggesting that the exciton and biexciton share the same electric dipole transition that is linearly oriented in the quantum dot. Thus, we assume that the exciton and the biexciton are both associated with the same quantum-dot-like emitter. For this SPE, laser power densities of less than 13 nW/$\mu m^2$ fully suppress the biexciton, as shown in Supplementary Fig. S15A.

To better understand the correlation between the strain-induced quantum confinement and the extraordinary enhancement of GaSe sub-bandgap emission (< 2.05 eV), we analyze the localized strain distribution on individual pillars by measuring the height profiles of GaSe flakes on pillars with atomic force microscopy (AFM). Fig. 4A shows an AFM image with a dimension of 3 $\mu m$ × 3 $\mu m$, where a multilayer GaSe flake (~ 30-nm-thick) on a pillar with a diameter of ~ 200 nm formed a tent with a diameter of ~ 2 $\mu m$ without any piercing. The classical Landau continuum model(43) for structural stress analysis (which has been commonly used to describe the nanoscale strain in nano-cubes,(1) nano-trenches,(44) and nano-bubbles(45)) is used to obtain the strain distribution across the pillar. More details about the strain simulation are shown in Supplementary Section 12 and Fig. S16. Fig. 4B illustrates the simulated position-dependent strain obtained from the AFM height profiles shown in Fig. 4A. The resolution of the strain simulation is set to be ~ 700 nm (yielding an 8 ×8 strain matrix) so that it is slightly coarser than the PL mapping resolution of ~ 400 nm.

Fig. 4C illustrates four representative PL spectra that are labeled in Fig. 4B, where PL spectra 1 (black) and 4 (blue) are collected at the edge of the pillar dome (corresponding to the minimum strain). In contrast, PL spectra 2 (red) and 3 (green) are collected at the pillar apex (corresponding to the maximum strain). Well-defined band edge emission $E_g$ at 2.02 eV (with relatively weak sub-bandgap emission $X_S$ at 1.69 eV) is observed at sites 1 and 4. At sites 2 and 3, several sub-bandgap peaks appear at ~ 1.69 eV with brightness about two orders of magnitude larger than those in spectra 1 and 4; in contrast, the band edge emission $X_{E_g}$ becomes much weaker. To specify, the peaks between 1.95 eV and 2.02 eV in PL spectra 2 (red) and 3 (green) arise from the indirect and direct excitons of GaSe.(14) Additional PL spectra from this pillar are shown in Supplementary Fig. S17. A similar trend is also observed for other pillars, as shown in Supplementary Figs. S18 and S19. In general, unlike the general co-presence of sub-bandgap and band edge emission on selenium clusters,(18) the sub-bandgap and band edge emissions on the pillar could be efficiently enhanced and suppressed, respectively.

Moreover, Fig. 4D illustrates the position-dependent integrated PL intensity ratio ($\frac{I_{X_S}}{I_{X_{E_g}}}$) map together with the calculated strain at different locations: the ratio of the sub-bandgap emission to the band edge emission ($\frac{I_{X_S}}{I_{X_{E_g}}}$) is enhanced by roughly two orders of magnitude on the pillar compared to that on the bare substrate area. A Pearson correlation analysis of the strain and PL intensity ratio ($\frac{I_{X_S}}{I_{X_{E_g}}}$) provides a quantitative evaluation of their correlation strength. Note that Pearson correlation is a measure of linear correlation between two sets of data that can be written as:

$$R = \frac{\sum(x_i-\underline{x})(y_i-\underline{y})}{\sqrt{(x_i-\underline{x})^2(y_i-\underline{y})^2}},$$

where R is the Pearson correlation coefficient, $x_i$ and $y_i$ are two sets of data, and $\underline{x}$ and $\underline{y}$ are their mean values, respectively. A Pearson correlation coefficient larger than 0.7 suggests a strong positive correlation between the two sets of data.(*20*) It has been demonstrated to effectively interpret the linear correlation between two different features in PL spectroscopies of 2D semiconductors.(*46–48*) Fig. 4E shows a strong positive Pearson correlation (R = 0.8) between the strain and the sub-bandgap emission intensities $I_{X_S}$; Fig. 4F further shows a strong Pearson correlation (R = 0.9) between the stain and the PL intensity ratio ($\frac{I_{X_S}}{I_{X_{E_g}}}$); a two-order enhancement of $I_{X_S}$ maps to a larger-than-0.2% biaxial tensile strain. However, $I_{X_{E_g}}$ has a weak negative correlation with the strain (R = -0.4) and $I_{X_S}$ (R = -0.45), and it vanishes when the biaxial tensile strain is larger than 0.2% (Supplementary Figs. S20C and D). In addition, the strain and the redshift of sub-bandgap emission from the band edge also show a strong positive correlation (R=0.731) (Supplementary Figs. S20A and B).

Previous reports of the photoluminescence from strained GaSe(*16, 49, 50*) have shown significant enhancement of the band edge emission. Optical absorption easurements(*49, 51, 52*) under a backscattering configuration have revealed that the optical absorption coefficient of GaSe with the light polarization perpendicular to the c axis ($\alpha_\perp$) is one order of magnitude larger than that parallel to the GaSe c axis ($\alpha\|$). Due to the tented structure of GaSe on the pillar, in the strained region, there is more contribution from $\alpha_\perp$ to the optical absorption, while in the unstrained region, the contribution from $\alpha_\perp$ is negligible. Increased optical absorption in the strained region could partially explain the extraordinary enhancement of sub-bandgap emission $I_{X_S}$ (as shown in Fig. 4E), but it does not explain the origin of the sub-bandgap emission $I_{X_S}$ nor the attenuated band edge emission $I_{X_{E_g}}$. The red shift of the band gap due to the localized biaxial tensile strain might explain these observations. Therefore, we perform band structure calculation of biaxially strained bulk GaSe for a qualitative comparison via the density functional theory (DFT) approach. Despite the common underestimation of the bandgap by DFT, it still reflects the correct physical trend. Our results show a bandgap redshift rate of -0.18 eV/% under biaxial tensile strain (Supplementary Fig. S21).

The DFT simulations also suggest that the valence band maximum (VBM) and conduction band minimum (CBM) show a blueshift rate of 0.086 eV/% and a redshift rate of -0.095 eV/%, respectively. This strain-induced opposing band bending behavior is defined as "Type-I funneling"(*53*), where the energy level of excited electrons descends towards the pillar apex with a larger strain while the energy level of holes ascends. As a result, the electrons and holes confined at the pillar apex have much higher concentrations than those at the flat region, which facilitates electron-hole recombination, leading to the enhancement of sub-bandgap emission and attenuation of band edge emission.

**Discussion**
we demonstrate the localization of quantum dot-like single-photon emitters in multilayer GaSe through strain control with SiO$_2$ nanopillar arrays. Ultimately, the results presented here provide a framework for stable, deterministic, and integrated quantum photonics. The GaSe SPEs arise from the exciton peak due to quantum confinement introduced by the nanoscale localized strain. The GaSe SPE purity is reduced with increasing excitation power density due to the formation of biexcitons. The incident laser power density required to excite an SPE varies at different sites due to the spatially varying confinement potentials. The local strain has a strong positive correlation with the sub-bandgap PL emission intensities and provides a two-order-of-magnitude enhancement in $\frac{I_{X_S}}{I_{X_{E_g}}}$. We attribute this increasing brightness of sub-bandgap emission to the strain-induced redshift of band gaps and the consequent "Type-I" exciton funneling. We envision that improved control over the localized charge concentration could help suppress biexciton PL and allow for bright, robust, and deterministic GaSe SPEs. We anticipate this demonstration of the deterministic localization of GaSe SPEs will stimulate future theoretical and experimental studies on SPEs in 2D materials and promote the development of quantum photonic devices and photonic quantum information technology.

**Materials and Methods**

**Synthesis of bulk GaSe crystal.** Bulk GaSe was synthesized by the chemical vapor transport (CVT) method.[9] A stoichiometric quantity of elements (molar ratio Ga: Se = 1:1, 1 g in total) and pure iodine (around 30 mg) were sealed into a quartz ampule under the pressure of < $10^{-5}$ Pa. Then, the ampule was put in a two-zone furnace and kept at 820 - 870 °C, with the precursor side placed in the high-temperature zone, for a week. Van der Waals GaSe single crystals were collected from the low-temperature zone when the growth finished.

**Fabrication of SiO$_2$ nanopillar arrays.**
1. A 8 mm × 8 mm silicon chip (cut by a dicing saw) with a 300-nm-thick thermal oxide layer and alignment markers was spin-coated with PMMA photoresist (MicroChem 950 A3) at 2000 rpm for 45 s, and a soft bake at 180 °C for 5 mins followed.
2. The SiO$_2$ nanopillar arrays were patterned onto the chip by electron beam lithography (EBL) via Nanometer Pattern Generation System (NPGS) on a Zeiss Supra 40 scanning electron microscope. Then the chip was developed in the isopropyl alcohol (IPA) / methyl isobutyl ketone (MIBK) mix solution (IPA: MIBK = 3:1) for 90 s, and a hard bake at 90 °C for 5 mins followed.
3. A 50-nm-chromium (Cr) protection layer was deposited (deposition rate: 0.5 Å/s) onto the chip via a CHA e-beam evaporator. A lift-off process in acetone was then performed to remove the Cr layer and keep only the developed region.
4. A reactive ion etching (RIE) process with fluoroform (CHF$_3$) / O$_2$ mixture gas was then performed on a Plasma-Therm 790 RIE etcher to etch a certain depth of SiO$_2$. A 9:1 (CHF$_3$) / O$_2$ and 40 Torr pressure would guarantee an etching rate of 70 nm/s on our machine.
5. After the RIE process, the residual Cr on the pillar was removed by immersing in the chromium etchant for 5 mins. Then the chip was rinsed with acetone, IPA, and DI water and treated with oxygen plasma to remove possible contaminants.

**Transfer of GaSe flake SiO$_2$ nanopillar arrays.** The GaSe flakes were first exfoliated onto a blank Si chip, and then a thin layer of poly-propylene carbonate (PPC) was spin-coated onto the chip. After curing at 60 °C for 5 mins, the PPC layer carrying the GaSe flakes was detached from the chip and transferred onto the SiO$_2$ nanopillar arrays under the microscope.

**Optical spectroscopy**. The cryo-PL and associated photon statistics measurements were performed in a home-built confocal PL microscope in a backscattering configuration. A Princeton Instruments Isoplane SCT-320 spectrograph with a Pixis 400BR Excelon camera and a grating turret with 150 g/mm, 600 g/mm and 2400 g/mm gratings was used to measure PL spectra with spectral resolution of: ~ 3.25 meV, ~ 300 $\mu$eV, and ~ 30 $\mu$eV, respectively. A 532 nm diode laser (Cobolt) was used for excitation. A 100x in-vacuum objective (Zeiss, NA = 0.85) was integrated in the Montana S100 closed-cycle cryostat. The PL mapping was controlled by 2-axis galvo scanning. The photon-antibunching measurements utilized a pair of large-area superconducting nanowire single-photon detectors (SNSPDs, Quantum Opus) and a Picoquant Hydraharp time-correlated single photon counting (TCSPC) system. A 90:10 non-polarizing beam splitter was used to allow for PL (10% coupling efficiency) and photon correlation functions (90% coupling efficiency) to be acquired in parallel.

**DFT simulations.** Plane-wave DFT calculations were carried out using the Vienna Ab initio Simulation Package[28,29] (VASP) with projector augmented wave (PAW) pseudopotentials[28,30,31] for electron-ion interactions, and the generalized gradient approximation (GGA) functional of Perdew, Burke and Ernzerhof(*54*) (PBE) for exchange-correlation interactions. Based on the bulk GaSe structure (Materials-project database(*55*)), the strain-free and biaxial-strained GaSe cell were optimized with a cutoff energy of 350 eV and 12×12×2 k-point samplings until the maximum force allowed on each atom was less than 0.001eV/Å. The total volume of the structure was fixed during geometry optimization to avoid the structural collapse of the 2D slabs with vacuum separations. The post-analysis of electronic band structures was carried out by using the VASPKIT package(*56*).

**Acknowledgments**
**Funding:** This material is based upon work supported by the National Science Foundation (NSF) under Grant No. (1945364). Work by X.L. was supported by the U.S. Department of Energy (DOE), Office of Science, Basic Energy Sciences (BES) under Award DE-SC0021064. The photoluminescence and photon statistics measurements were performed at the Center for Nanophase Materials Sciences (CNMS), which is a US Department of Energy Office of Science User Facility. W.L. acknowledges Dr. Y.Y. Pai from Oak Ridge National Laboratory for his help with experimental automation. X.L., A.S. and A.K.S. acknowledge the membership of the Photonics Center at Boston University. The computational work is performed using Shared Computing Cluster at Boston University.

**Author contributions:**
Conceptualization: WL, XL
Methodology: WL, AP, BL, QT, HG, ZC, XL
Investigation: WL, XL, AP, BL
Visualization: WL, XL
Supervision: XL, AP, BL
Writing—original draft: WL, XL, AP, BL
Writing—review & editing: WL, XL, AP, BL, QT, HG, ZC, AS, LL, AKS

**Competing interests:** Authors declare that they have no competing interests

**Data and materials availability:** All data are available in the main text or the supplementary materials.

**Figures and Tables**

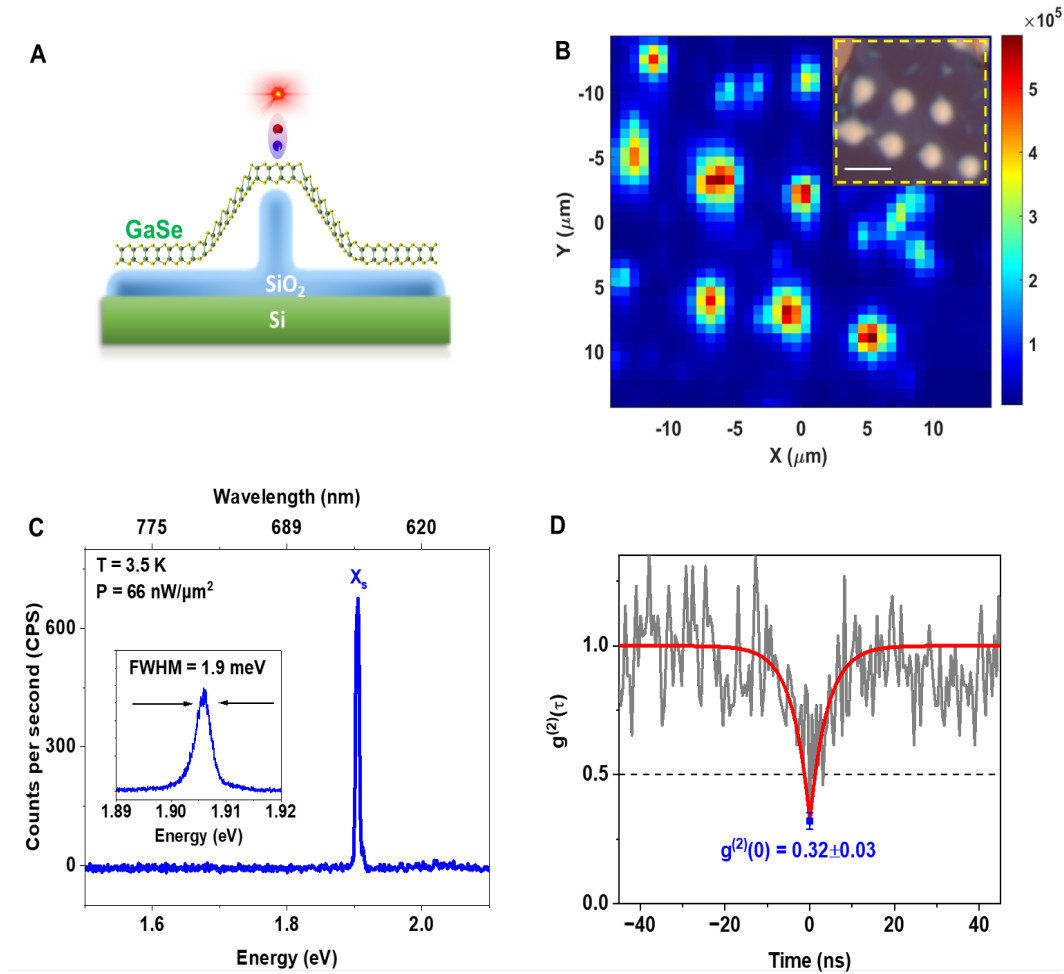

**Fig. 1 | GaSe single photon emitters on nanopillar arrays. (A)** Illustration of GaSe thin film tented on the pillar; for simplification, we used monolayer GaSe instead of multilayer GaSe in the illustration. **(B)** PL intensity map integrated across wavelengths of 600-800 nm for a representative array of pillars measured at T = 3.5 K, with an incident laser power density of 2 $\mu W/\mu m^2$; the inset optical image (scale bar, 5 $\mu$m) shows an exfoliated GaSe flake transferred onto SiO$_2$ nanopillar arrays. **(C)** A representative PL spectrum of GaSe on pillar apex with an incident laser power density of 66 nW/$\mu m^2$) measured at T = 3.5 K. The single photon emission is labeled as $X_S$. **(D)** Photon-antibunching measured for $X_S$ acquired using the same measurement conditions and a narrow bandpass filter (650 nm, full width at half maximum (FWHM) = 10 nm); the fitted second-order correlation function shows $g^{(2)}(0) = 0.32\pm0.02$.

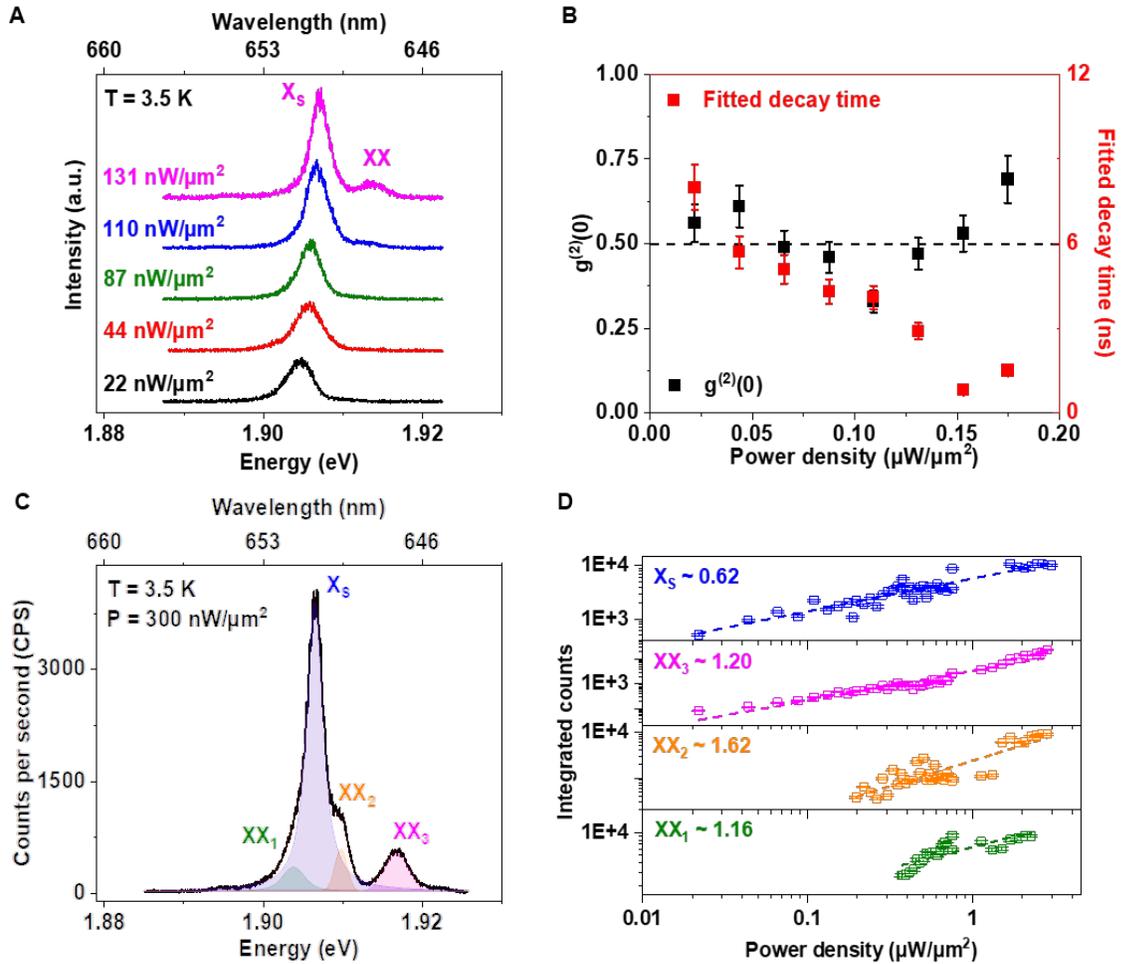

**Fig. 2 | Power density-dependent PL spectra and photon statistics measurements of the same GaSe SPE at ~1.90 eV (~ 652 nm) (shown in Fig. 1C) acquired at T = 3.5 K. (A)** Selected power density-dependent PL spectra of the SPE under relatively low excitation power density (< 150 nW/$\mu m^2$). **(B)** Photon antibunching measurements of this SPE under different incident laser power densities. **(C)** Deconvolution of a PL spectrum of the SPE excited under a relatively high excitation power density of ~ 300 nW/$\mu m^2$ that consists of three biexcitons ($XX_1$, $XX_2$, and $XX_3$) and one exciton ($X_S$) feature. **(D)** The integrated counts of the three biexcitons and the exciton as a function of incident laser power density.

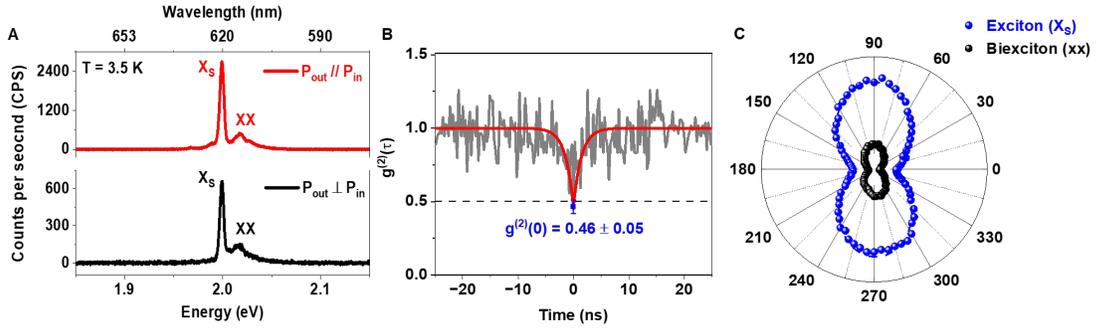

**Fig. 3 | Characterization of a GaSe SPE at 1.99 eV (~ 620 nm) with a horizontally polarized excitation at T = 3.5 K.** The incident laser power density is 110 nW/$\mu m^2$ and a narrow bandpass filter (622 nm, FWHM = 10 nm) was used for antibunching measurements. **(A)** Parallel- and cross-polarized PL spectra. **(B)** Second-order correlation function for the same SPE with the unpolarized collection. **(C)** Integrated polarization-dependent PL intensities from the exciton at 1.99 eV (~ 620 nm) and biexciton at 2.02 eV (~ 615 nm), respectively.

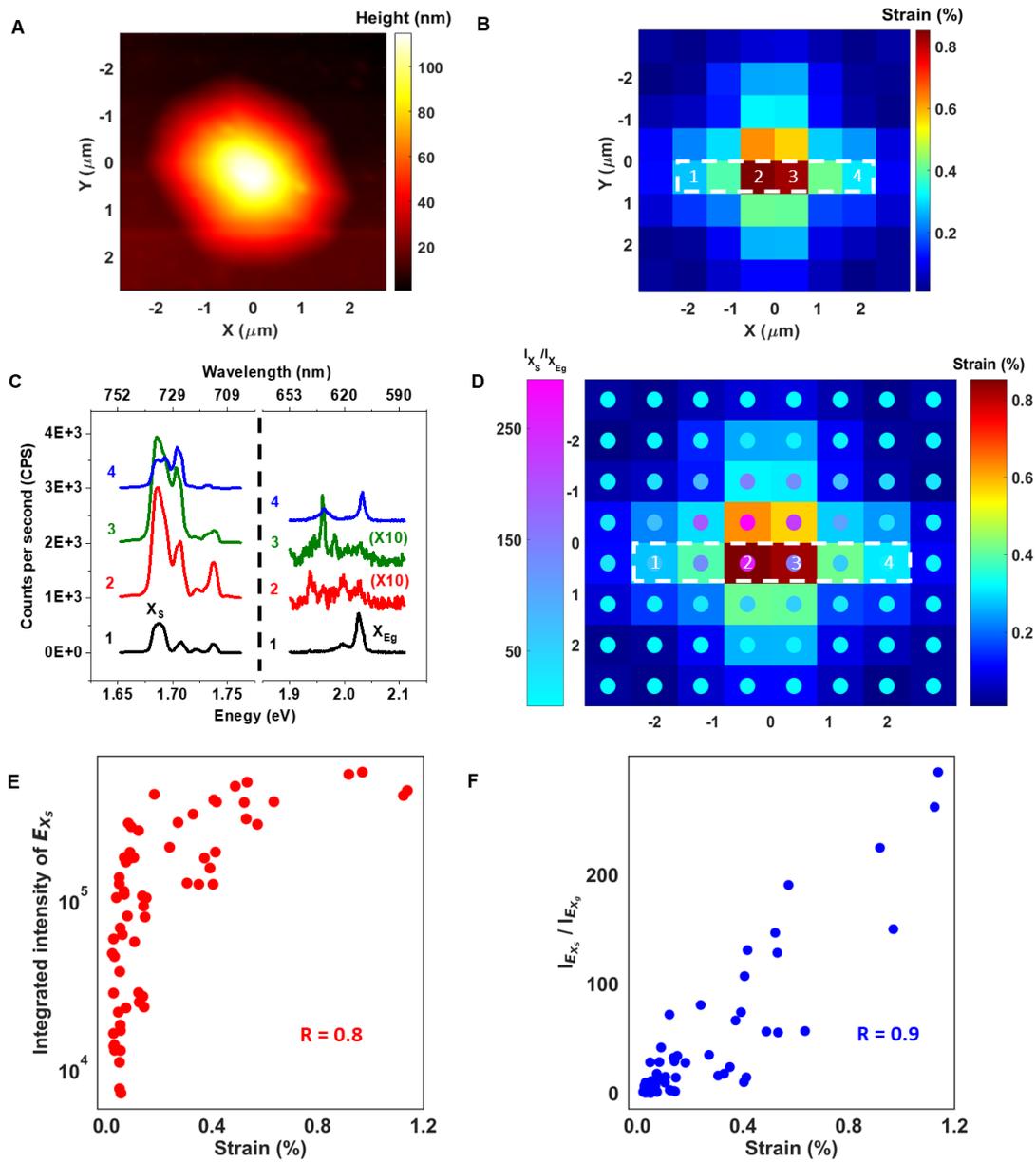

**Fig. 4 | Correlation of strain distribution analysis with PL mapping results at T = 3.5 K. (A)** AFM height profile of pillar 1 (marked in Supplementary Figs. S5 and S6). **(B)** Strain map obtained from the AFM height profile in Fig 4**(A)**: each pixel covers an area of 0.7 $\mu m \times$ 0.7 $\mu m$, and the corresponding color represents the averaged strain (biaxial tensile strain). **(C)** Four PL spectra collected at the numbered positions as shown in Fig 4**(B)**. **(D)** Correlation map of strain (squares, adapted from Fig.4**(B)**) and the integrated PL intensity ratios (circles) between the sub-bandgap (strain-confined exciton) and band edge emission $I_{X_S/X_{Eg}}$ plotted on a 5.6×5.6 $\mu m$ grid. **(E)**, Pearson correlation analysis of strain and the integrated sub-bandgap intensity $I_{X_S}$. **(F)** Pearson correlation analysis of strain and the integrated intensity ratios of $I_{X_S/X_{Eg}}$.

**Supplementary Materials**

Fig. S1. Crystal structure and characterization of GaSe crystal.
Fig. S2. Schematics of sample preparation.
Fig. S3. SEM Images of the pillar topography.
Fig. S4. 3D AFM images of the pillar topography.
Fig. S5. AFM image of the sample shown in the inset figure of Fig. 1(B) of the main text.
Fig. S6. Full PL mapping image (at T = 3.5 K) of Fig. 1(B) of the main text.
Fig. S7. Histogram of the emission wavelengths of PL mapping results of Fig. 1(B) of the main text.
Fig. S8. Characterization of SPE and non-SPEs at T = 3.5 K.
Fig. S9. Characterization of emitters at T = 3.5 K.
Fig. S10. Characterizations of emitters with reduced purity at T = 3.5 K.
Fig. S11. Spectral wandering of the GaSe SPE at ~ 652 nm shown in Fig. 2 of the main text.
Fig. S12. Incident power-dependent antibunching and PL spectra analysis.
Fig. S13. The temperature-dependence of a GaSe SPE at ~ 693 nm.
Fig. S14. Photon-antibunching measurements (at T = 3.5 K) of the GaSe SPE at ~ 693 nm are shown in Fig. S13 with different incident power intensities.
Fig. S15. Power-dependent PL measurements of the SPE at ~ 621 nm are shown in Fig. 3 of the main text.
Fig. S16. Simulation of strain distribution based on pillars 1 – 3 labeled in the AFM image shown in Fig. S5.
Fig. S17. PL spectra (at T = 3.5 K) of different rows of pillar 1 presented in Fig. 4 of the main text.
Fig. S18. Analysis of the sub-bandgap emission, the strain-induced redshift of emission energy, and intensity ratio of $\frac{I_{X_S}}{I_{E_g}}$ of pillars 2 – 4.
Fig. S19. Analysis of the sub-bandgap emission, the strain-induced redshift of emission energy, and intensity ratio $\frac{I_{X_S}}{I_{E_g}}$ of pillars 5 – 7.
Fig. S20. Correlation analysis between strain simulation results and analysis of PL spectra.
Fig. S21. DFT simulation results of strained GaSe.

# Supplementary Materials for

## Deterministic Localization of Strain-induced Single-photon Emitters in Multilayer GaSe

Weijun Luo, Alexander Puretzky, Benjamin Lawrie, Qishuo Tan, Hongze Gao, Zhuofa Chen, Alexander Sergienko, Anna Swan, Liangbo Liang, Xi Ling

*Corresponding author. Email: xiling@bu.edu; puretzkya@ornl.gov; lawriebj@ornl.gov

**This PDF file includes:**

# Supplementary Text

Section 1. Schematics of GaSe crystallographic structure

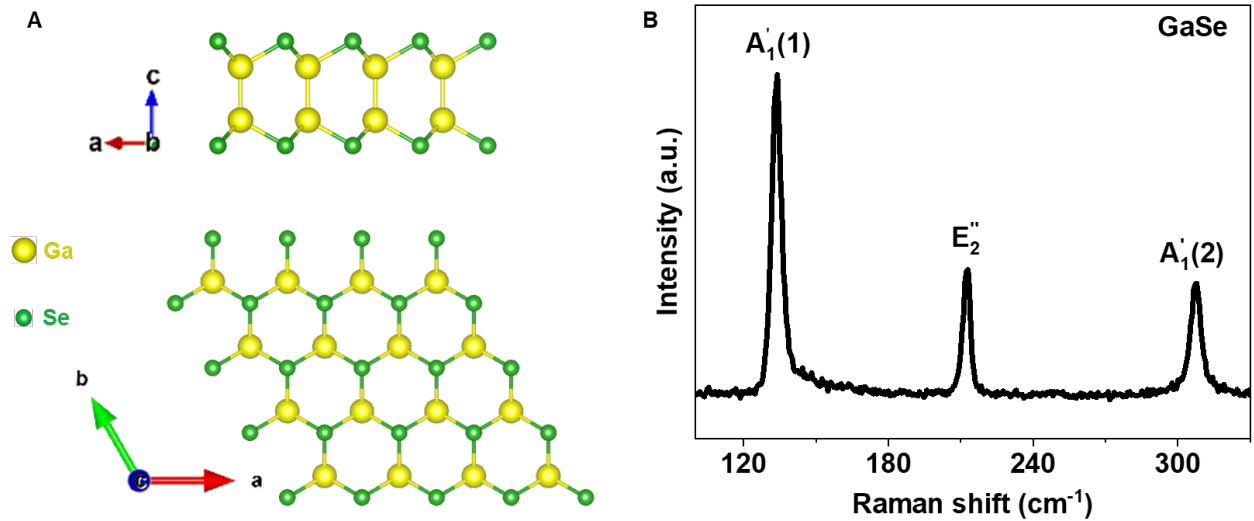

**Supplementary Fig. S1 | Crystal structure and characterization of GaSe crystal. (A)** Side- and top-view illustration of the crystallographic structure of monolayer GaSe. **(B)** Raman spectra of synthesized GaSe bulk crystal.

Note that GaSe has three polytypes: $\beta$, $\varepsilon$, and $\gamma$. $\varepsilon$-type GaSe with a hexagonal structure and an indirect bandgap of ~2.05 eV is usually preferentially grown by vapor phase transport and the Bridgman technique(*1*). In bulk $\varepsilon$-GaSe, the stacking follows an AB sequence and belongs to the non-centrosymmetric space group P$\underline{6}$m2 (No.187) and point group of $D_{3h}^1$. Fig. S1(A) shows the side view and top view of a $\varepsilon$-GaSe layer: in each layer, each Ga atom is covalently bonded with three Se atoms and one Ga atom; these layers are held together by weak van der Waals forces with a layer distance of ~ 0.8 nm(*2*). Fig. S1(B) shows the Raman spectrum of the synthesized bulk GaSe, where $A_1'(1)$, $E_2''$ and $A_1'(2)$ modes are at $134\ cm^{-1}$, $213\ cm^{-1}$, and $308\ cm^{-1}$, respectively, matching the previous reported values.(*3*, *4*)

Section 2. Schematics of sample preparation

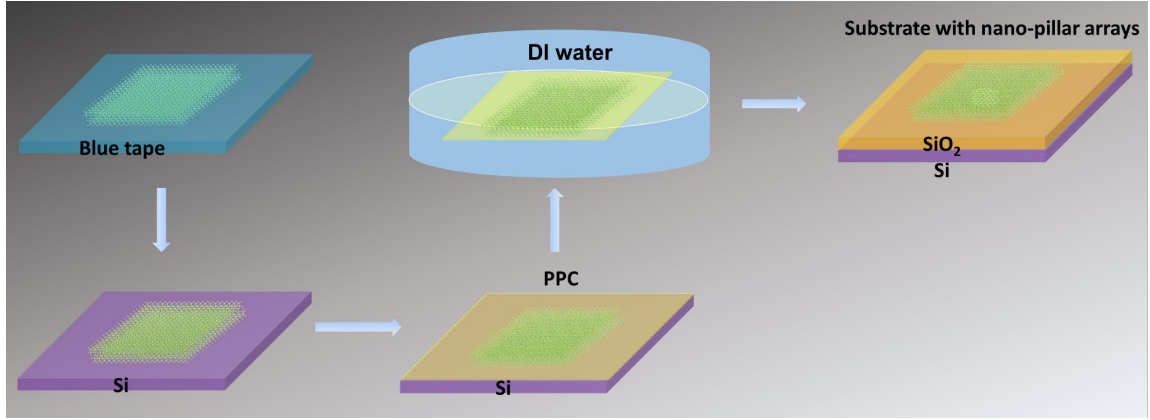

**Supplementary Fig. S2 | Schematics of sample preparation**.

The mechanically exfoliated GaSe flakes were picked up by the spin-coated Polypropylene carbonate (PPC) thin layer and transferred to Si chips with pre-patterned $SiO_2$ nanopillar arrays. The PPC layer was removed by acetone and then rinsed with IPA and deionized water.

**GaSe exfoliation and transfer:**

1) Multilayer GaSe flakes were first mechanically exfoliated by Nitto SPV 224R blue tape from bulk GaSe single crystal and transferred onto a Si chip with a 300-nm-thick thermal oxide layer.
2) Polypropylene carbonate (PPC) thin layer (1:10 dissolved in anisole) was spin-coated (2000 rpm, 30 sec) on the Si chip and cured at 80 ℃ for 5 minutes.
3) The PPC film carrying GaSe flakes was peeled off in DI water.
4) After fast baking at 40 ℃ for 1 minute, the GaSe flakes were immediately transferred onto Si chips with pre-patterned $SiO_2$ nanopillar arrays.
5) A post-annealing step (100 ℃ for 1 hour) was followed to enhance the adhesion between GaSe flakes and pre-patterned substrate.

**The fabrication process of the nanopillar arrays**

(1) Nanopillar (150 nm, 200 nm, and 250 nm in diameter) arrays were patterned on PMMA (MicroChem 950 A3) photoresist using E-beam lithography.

(2) A 40-nm-thick Cr film (sacrificial layer) was deposited by e-beam evaporation on the photoresist after development in MIBK: IPA (1:3) solution for 100 sec.

(3) After lift-off in acetone for 5 mins, the Si chip was rinsed with IPA and DI water and then dried.

(4) A 2-minute plasma dry etching by a $CHF_3$ and $O_2$ (45 sccm: 5 sccm, 40 mTorr) mixture gas was performed to etch down $SiO_2$ isotropically for 110 nm.

(5) Residual Cr on the Si chip was then removed by Cr etchant.

Section 3. SEM and AFM images of pillar topography.

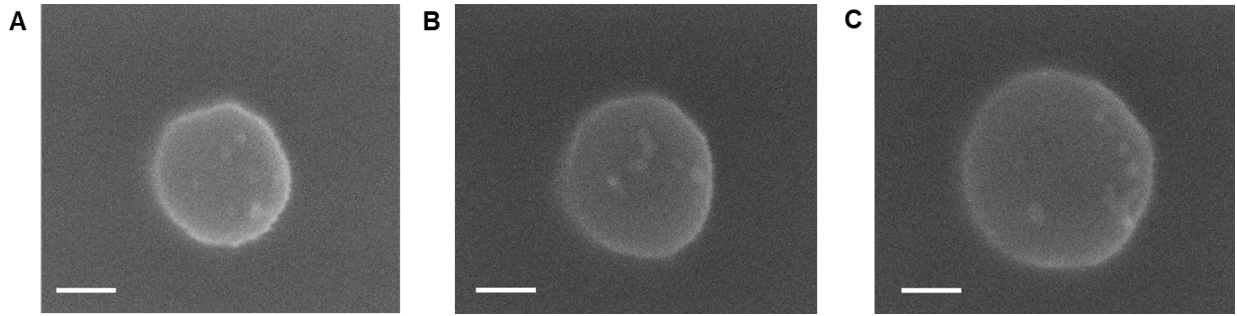

**Supplementary Fig. S3 | SEM Images of the pillar topography.** The scale bar is 100 nm. **(A)** pillar diameter of ~ 150 nm. **(B)** pillar diameter of ~ 200 nm. **(C)** and pillar diameter of ~ 250 nm. The scanning electron microscopy (SEM) images show that circle-shaped $SiO_2$ nanopillar arrays have three diameters of ~ 150 nm, ~ 200 nm, and ~ 250 nm, respectively.

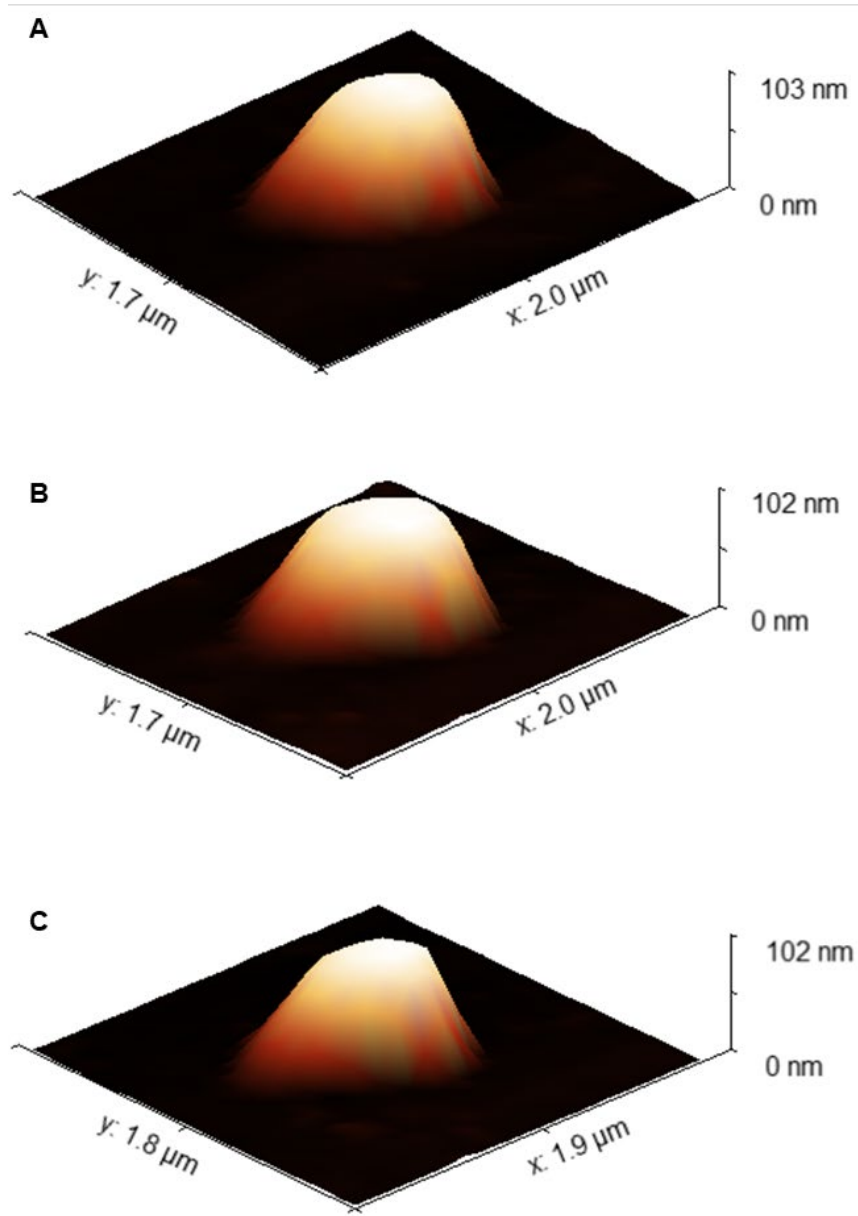

**Supplementary Fig. S4 | 3D AFM images of the pillar topography. (A)** pillar diameter of ~ 150 nm. **(B)** pillar diameter of ~ 200 nm. **(C)** and pillar diameter of ~ 250 nm; all the pillars have an average height of ~ 100 nm.

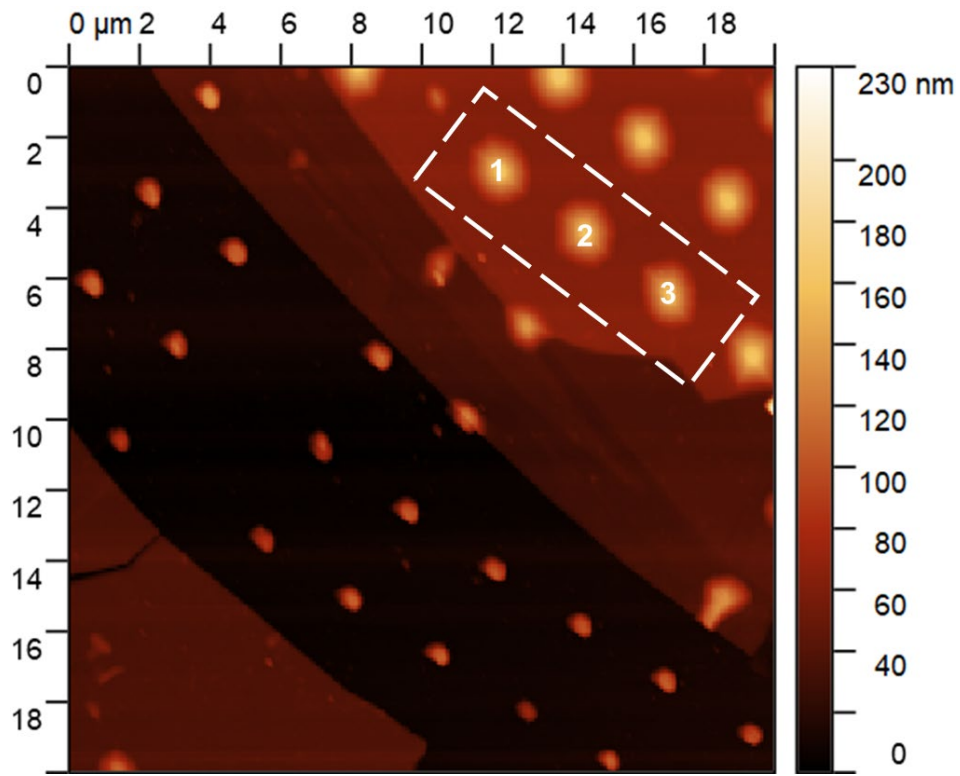

**Supplementary Fig. S5 | AFM image of the sample shown in the inset figure of Fig. 1(B) of the main text.**

Section 4. Analysis of the PL mapping results at T = 3.5 K.

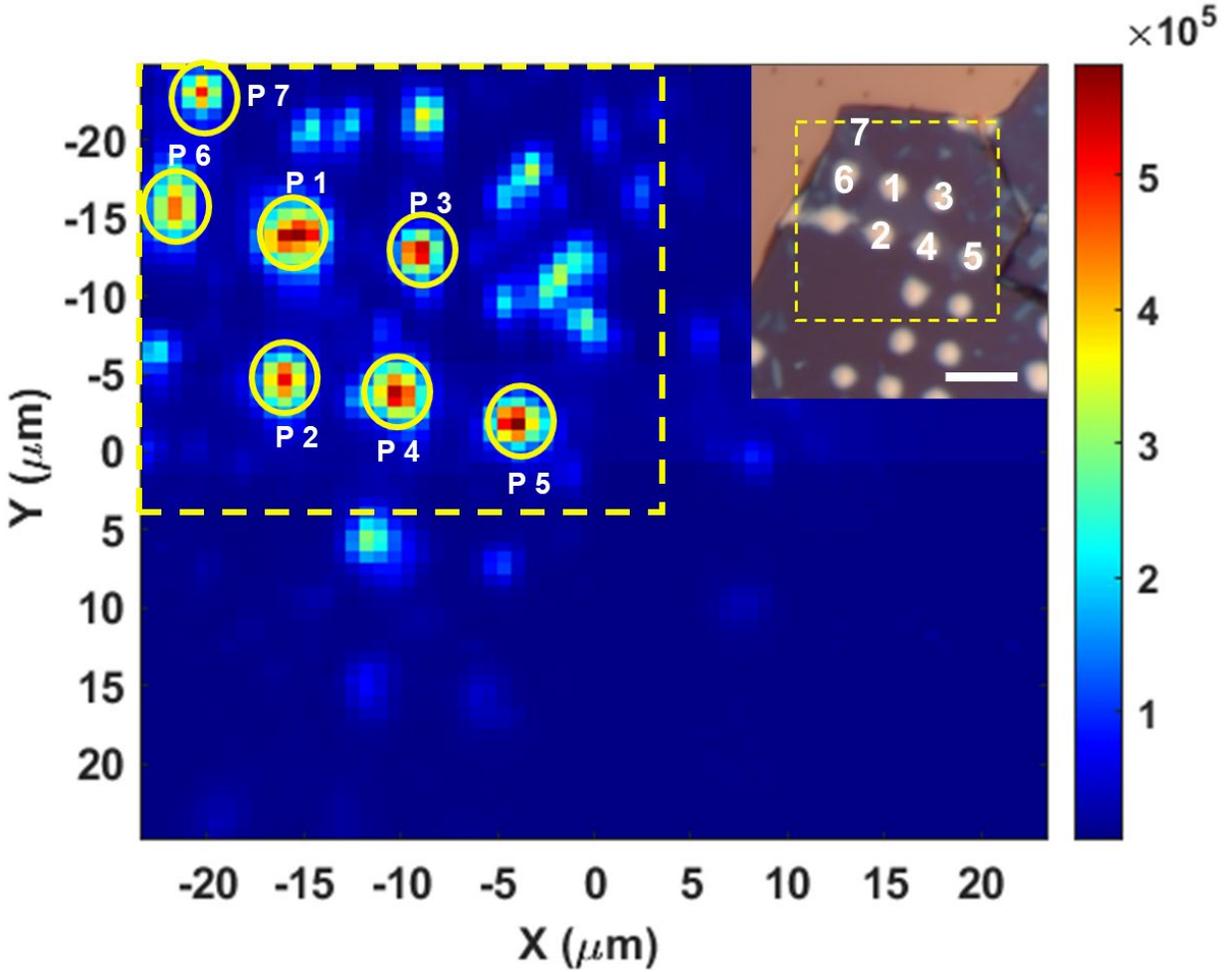

**Supplementary Fig. S6 | Full PL mapping image (at T = 3.5 K) of Fig. 1(B) of the main text.** The PL image shows the integrated intensities between 600 – 800 nm of a GaSe flake shown in the inset optical image (scale bar, 5 $\mu$m).

Section 5. Statistics of the emission wavelengths of PL mapping results of Fig. 1(B) of the main text.

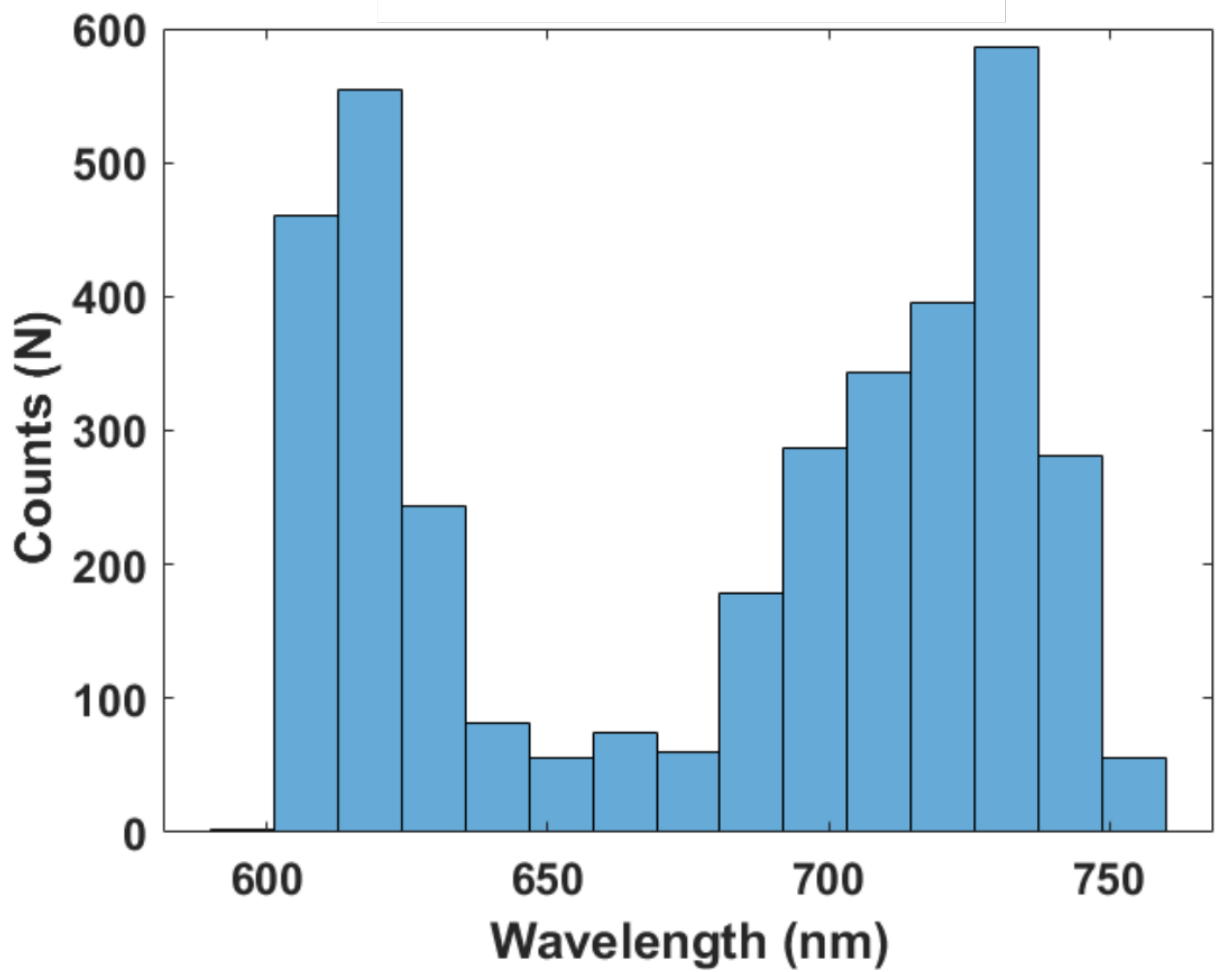

**Supplementary Fig. S7 | Histogram of the emission wavelengths of PL mapping results of Fig. 1(B) of the main text.**

Section 6. Antibunching fitting

The photon-antibunching results could be fitted by the second-order correlation function (intensity auto-correlation)(5):

$$g^{(2)}(\tau) = \frac{\langle I(\tau)I(t+\tau)\rangle}{\langle I(\tau)\rangle^2}, \quad (1)$$

where $I$ represent the emission intensity, and $\tau$ is the time delay. In practice, the intensity autocorrelation function could be written as:

$$g^{(2)}(\tau) = 1 - a * exp^{-\frac{|\tau|}{b}}, \quad (2)$$

where (1-a) and b are the photon-antibunching value at the zero-time delay ($\tau = 0$, $g^{(2)}(0)$) and antibunching decay time, respectively.

However, at higher excitation power density, the $g^{(2)}$ function superimposes a bunching $g^{(2)}(\tau) > 1$ feature at certain delays to the antibunching curve. Such correlation profile calls for a third metastable energy level that prevents subsequent emission of a photon for a certain time(6–8). The intensity autocorrelation function could be rewritten as:

$$g^{(2)}(\tau) = 1 - \beta * exp(-\gamma_1 \tau) + (\beta-1) * exp(-\gamma_2 \tau), \quad (3)$$

Section 7. Characterizations of SPEs.

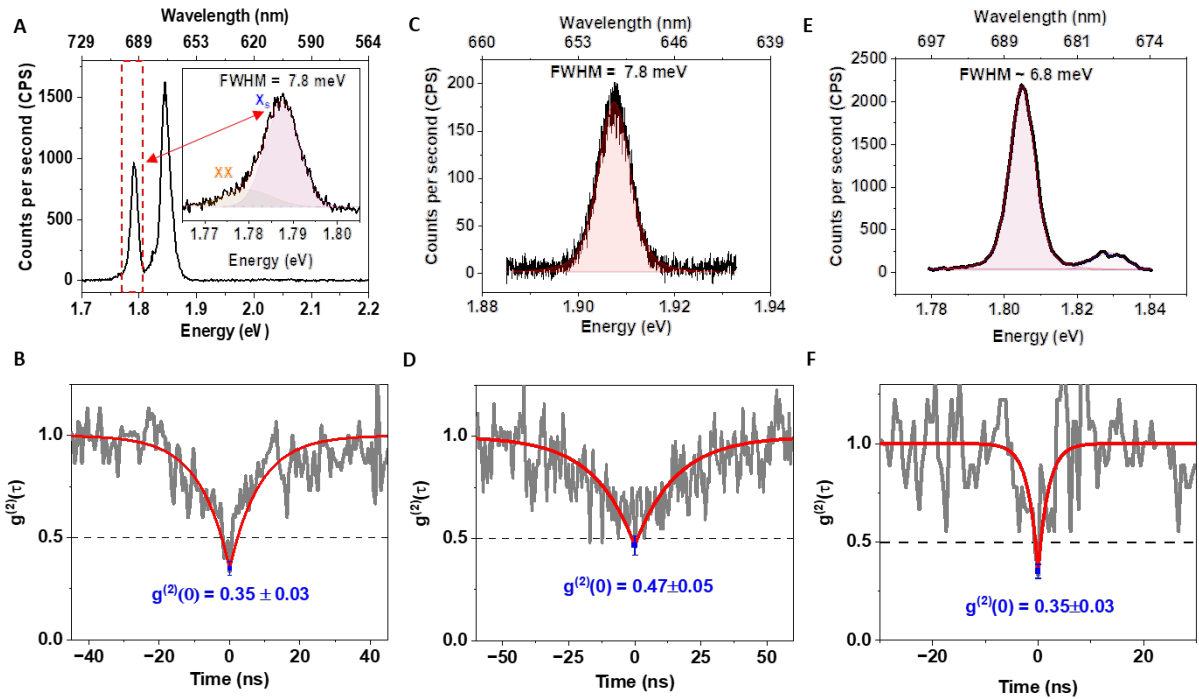

**Supplementary Fig. S8 | Characterization of SPE and non-SPEs at T = 3.5 K. (A)/(B)** PL spectra / photon-statistics measurements of a GaSe SPE at ~ 690 nm. **(C)/(D)** PL spectra / photon-statistics measurements of a GaSe SPE emitter at ~ 650 nm. **(E)/(F)** PL spectra / photon-statistics measurements of a GaSe SPE at ~ 688 nm.

A representative PL spectrum collected on a pillar, shown in Fig. S8(A), exhibits two isolated peaks with energies of ~ 1.79 eV and ~ 1.85 eV, about ~ 0.2 – 0.25 eV below the well-studied GaSe band edge emission at 2.05 eV[5,11]. Note that the emission energies of GaSe SPEs here are consistent with the previous work on GaSe SPEs associated with randomly distributed selenium clusters(*9*). Because each pillar introduces a continuous strain gradient over a sub-diffraction limited area, multiple peaks associated with different potential wells are observed for many pillars, particularly at higher excitation powers. Each peak exhibits different photon statistics because each peak is associated with a different confinement potential. The inset image in Fig. S8(A) shows the finer structure of the peak at ~ 1.79 eV together with a two-mode Gaussian fit that includes an intense narrow peak (~ 1.787 eV, FWHM ~ 8.0 meV) and a weak broad peak (~ 1.777 eV, FWHM ~ 17 meV). The higher energy peak at 1.787 eV was identified as an exciton, and the weak lower energy peak at 1.777 eV was identified as a biexciton[5,12] with a binding energy of $|\Delta E|$ ~ 10 meV (rather than a phonon sideband). A 690 nm bandpass filter with a FWHM of 10 nm and peak transmittance > 70% was used for the photon correlation function measurements to separate the two peaks shown in Fig. S8(A). The peak at ~ 1.85 eV does not exhibit measurable antibunching. Some of the measured GaSe SPEs show symmetric line shapes suggesting only one exciton peak, whereas others exhibit an asymmetry like that observed in Fig. S8(A), consistent with a superposition of exciton and biexciton PL.

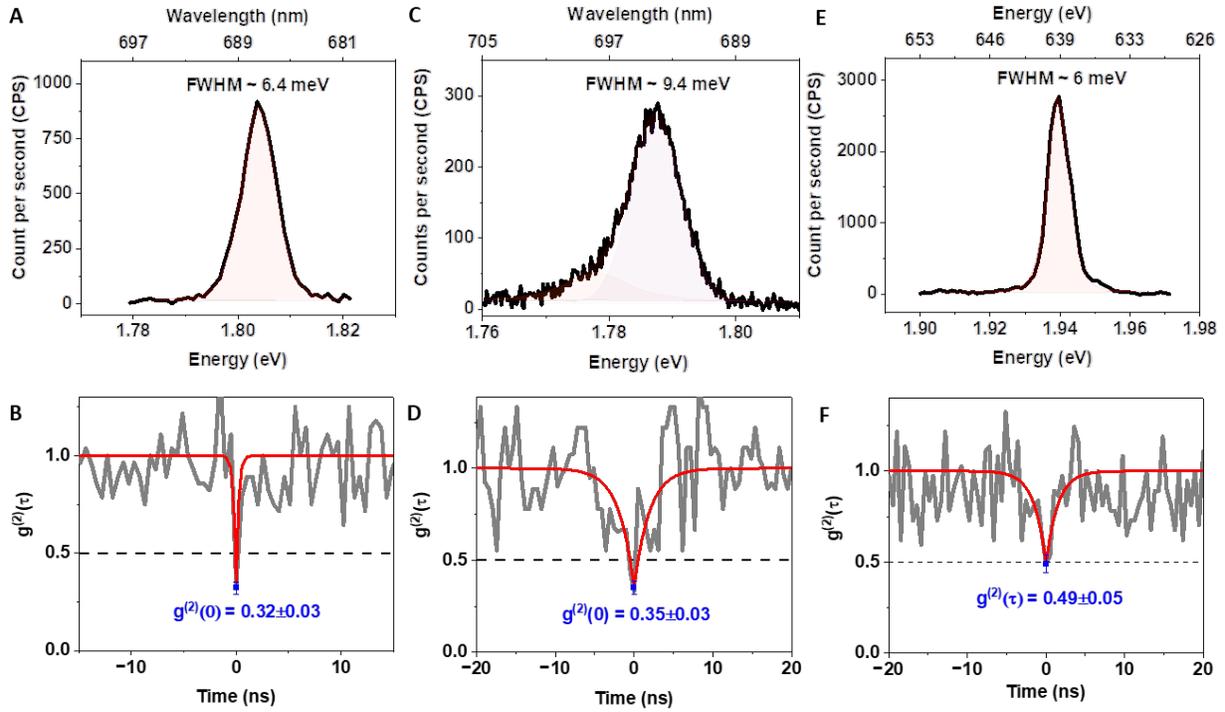

**Supplementary Fig. S9 | Characterization of emitters at T = 3.5 K. (A)/(B)** PL spectra / photon-statistics measurements of a GaSe SPE at ~ 688 nm. **(C)/(D)** PL spectra / photon-statistics measurements of a GaSe SPE at ~ 694 nm. **(E)/(F)** PL spectra / photon-statistics measurements of a GaSe SPE at ~ 640 nm.

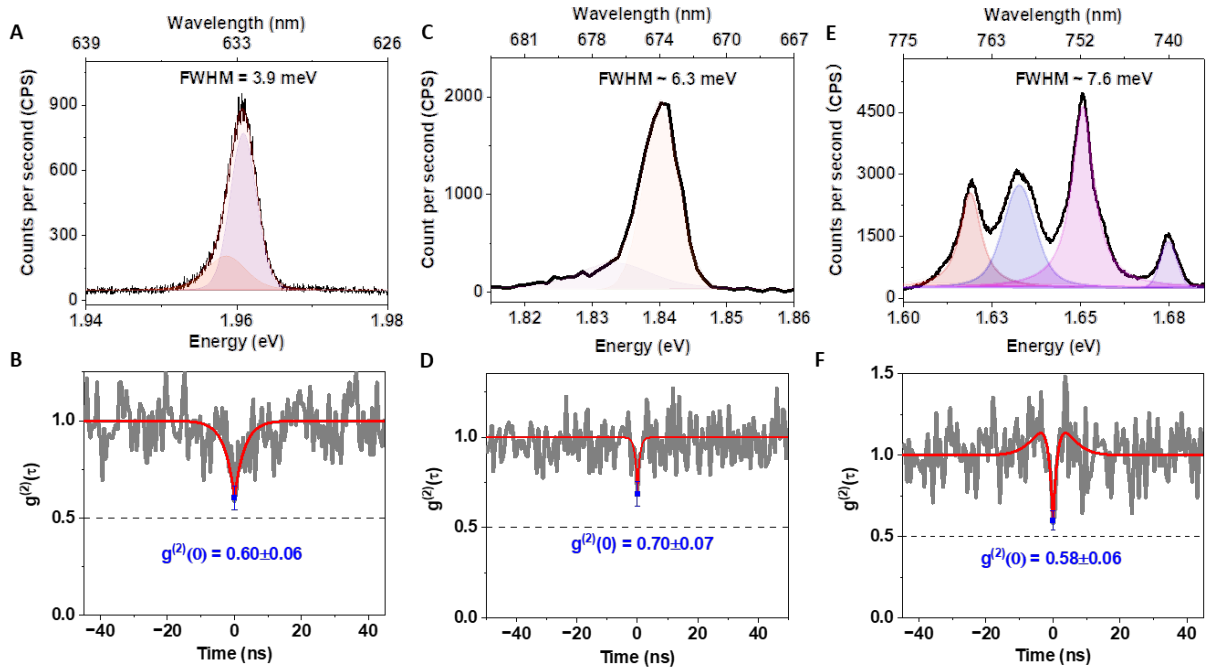

**Supplementary Fig. S10 | Characterizations of emitters with reduced purity at T = 3.5 K.** **(A)/(B)** PL spectra / photon-statistics measurements of an emitter at ~ 640 nm. **(C)/(D)** PL spectra / photon-statistics measurements of an emitter at ~ 673 nm. **(E)/(F)** PL spectra / photon-statistics measurements of an emitter at ~ 750 nm.

Section 8. The spectral wandering of the GaSe SPE at ~ 652 nm shown in Fig. 2 of the main text with different incident power densities

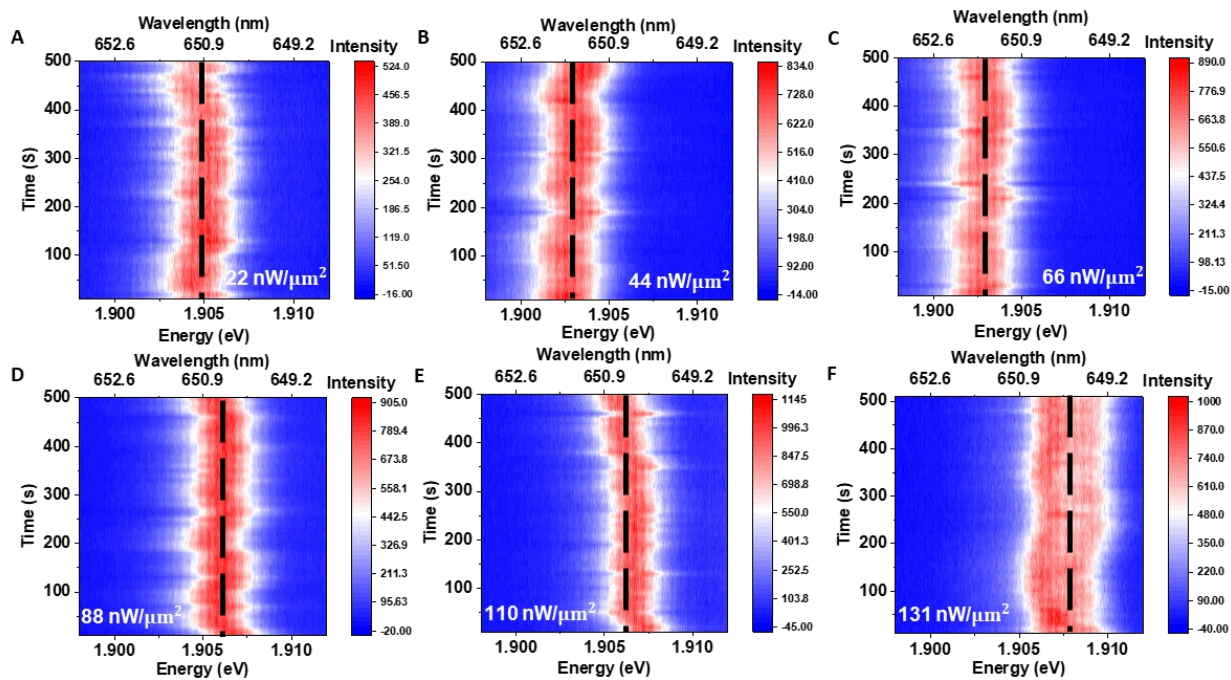

**Supplementary Fig. S11 | Spectral wandering of the GaSe SPE at ~ 652 nm shown in Fig. 2 of the main text.**

Section 9. Photon-antibunching measurements of the GaSe SPE at ~ 652 nm shown in Fig. 2 of the main text with different incident power densities

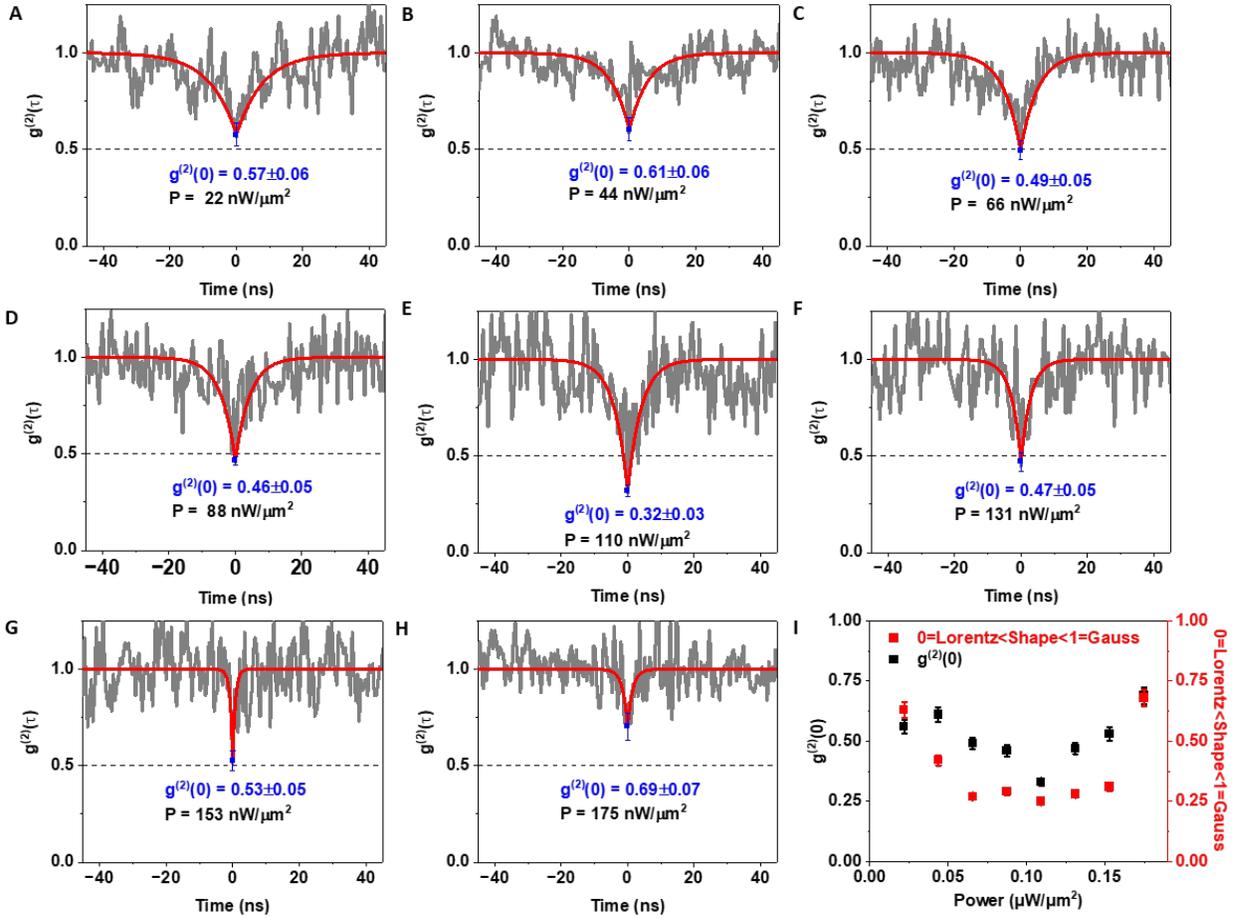

**Supplementary Fig. S12 | Incident power-dependent antibunching and PL spectra analysis. (A)-(H)** Photon-antibunching measurements (at T = 3.5 K) of the GaSe SPE at ~ 652 nm shown in Fig. 2 of the main text with different incident power densities. **(I)** $g^{(2)}(\tau = 0)$ and weight factor $\nu$ as functions of incident power densities.

To specify, the PL spectrum of an ideal SPE is with a Lorentzian line shape, corresponding to an exponentially decaying excited state with a finite life time(*10*). However, the PL spectral linewidth is affected by various environmental fluctuations such as heat(*11*), strain(*12*), and doping(*13*, *14*), which often results in inhomogeneous linewidth broadening that is described by the Gaussian function(*15*). Therefore, in practice, the Voigt function that results from the convolution between the Gaussian and Lorentzian function is more often adopted for the spectral line fitting of SPEs(*16*,

*17*). The Voigt function could be expressed as a weighted sum of Lorentzian and a Gaussian functions as(*18, 19*):

$$V(x) \approx (1-\nu)L_1(x) + (\nu)G_1(x), (1)$$

Where $G_1(x)$ and $L_1(x)$ are a Gaussian and a Lorentzian functions, and $0 \leq \nu \leq 1$. When $\nu = 1$, $V(x)$ is a Gaussian function; when $\nu = 0$, $V(x)$ is a Lorentzian function. Hence, the factor $\nu$ evaluates the weights of Gaussian and Lorentzian components. For the power-dependent PL spectra of the GaSe SPE at ~ 652 nm, the values were extracted and visualized in Fig. S12(I): the trend of weight factor $\nu$ matches that of $g^{(2)}(0)$, implying that the bottleneck limits the improvement of $g^{(2)}(0)$ values is the weight factor $\nu$. In other words, simply decreasing incident power densities do not monotonically improve the $g^{(2)}(0)$ values.

Section 10. Photon-antibunching measurements and PL spectra of the GaSe SPE at ~ 693 nm with different incident power densities.

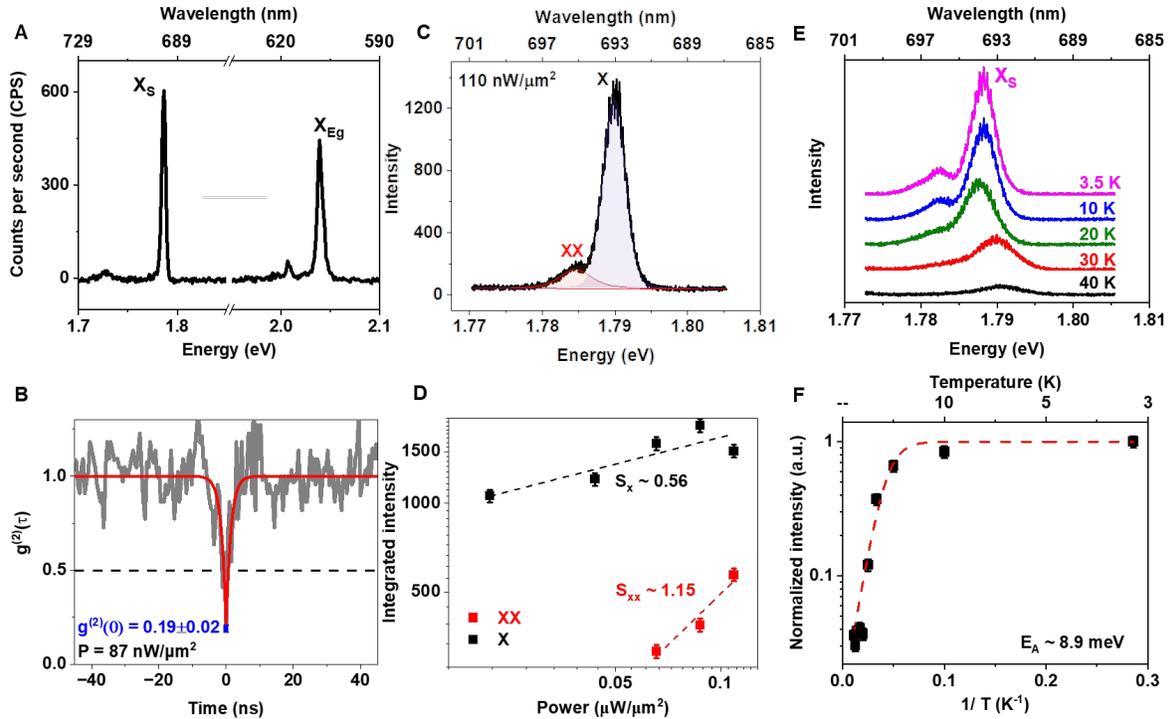

**Supplementary Fig. S13 | The temperature-dependence of a GaSe SPE at ~ 693 nm. (A)** PL spectrum (collected with 150 g/mm, grating and incident power density of 87 nW/$\mu m^2$. **(B)** the photon statistics measurement of this ~ 693 nm emitter, where a 692 nm narrow bandpass filter with an FWHM of 10 nm and peak transmittance > 90% was used. The minimum value of the recorded second-order correlation function $g^{(2)}(\tau = 0)$ is ~ 0.19. **(C)** PL spectrum (collected with 2400 g/mm) of this GaSe SPE at T = 3.5 K. **(D)** The incident power-dependent integrated intensities of exciton and biexciton of this ~ 693 nm emitter were plotted in a double-logarithmic plot. **(E)** PL spectra (collected with 2400 g/mm) of this GaSe SPE at different temperatures. **(F)** The temperature-dependent integrated intensities of this 693 nm emitter are plotted as a function of the reciprocal of temperature.

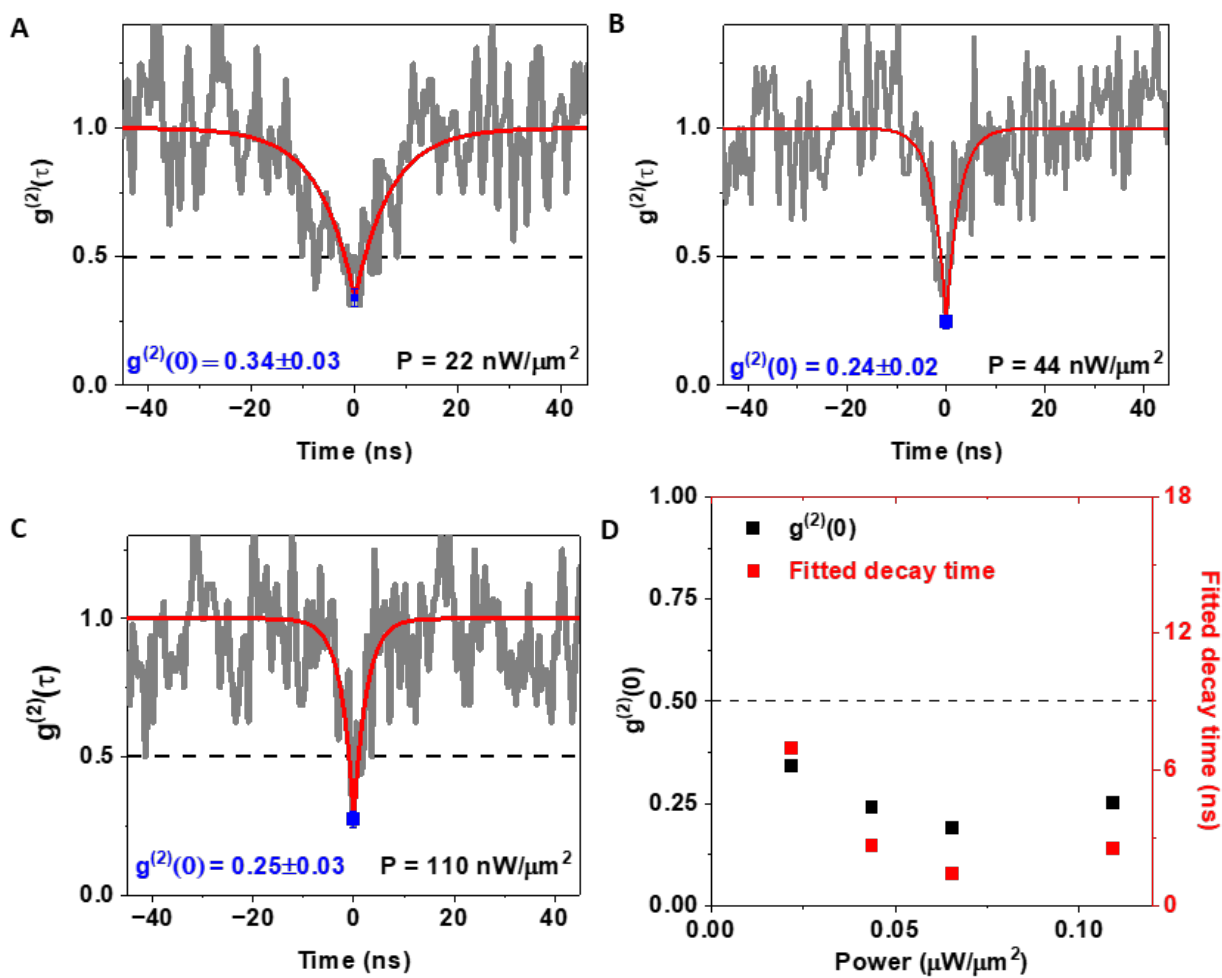

**Supplementary Fig. S14 | Photon-antibunching measurements (at T = 3.5 K) of the GaSe SPE at ~ 693 nm are shown in Fig. S13 with different incident power intensities. (A)** 22 nW/$\mu m^2$. **(B)** 44 nW/$\mu m^2$. **(C)** 110 nW/$\mu m^2$. **(D)** Photon antibunching measurement results of this ~ 693 nm emitter under different incident power and the corresponding fitted decay time are also integrated.

We also investigate the temperature-dependent performance of GaSe SPEs on the pillar. Fig. S13(A) shows a PL spectrum collected on a pillar: a sharp resonance peak at ~ 1.79 eV (~ 693 nm) was identified to be an exciton feature at a low incident power density of 87 nW/$\mu m^2$. Then the photon-statistics measurements at different incident powers were carried out using a 692 nm narrow bandpass filter with an FWHM of 10 nm and peak transmittance > 90%. Fig. S13(B) shows that at the incident power density of 87 nW/$\mu m^2$, a $g^2(0)$ value of ~ 0.15 is obtained. Fig. S13(C) shows the high-resolution PL spectra acquired with a 2400 g/mm grating of this ~ 1.79 eV SPE at different temperatures: at 3.5 K, this SPE shows an exciton feature at ~ 1.788 eV and a biexciton feature at ~ 1.782 eV ($|\Delta E| \sim 6$ meV). Fig. S13(E) illustrates that with increasing temperature, the biexciton feature first vanishes at ~ 20 K, while the exciton peak decreases by 90% at 40 K and quenches after 90 K. Note the quenching of emission could be explained by electron-hole recombination through the increasing number of nonradiative centers activated at elevated temperatures(*20*). The relationship between the integrated intensities and temperatures follows the Arrhenius equation(*21*):

$$I^{PL}(T) = \frac{I^{PL}(\to 0K)}{1+Cexp(-\frac{E_A}{kT})},$$

where C is a constant, $E_A$ is the thermal-quenching activation energy, and k is Boltzmann's constant. Fig. S13(F) shows the results Arrhenius plot of this ~ 1.79 eV SPE and the corresponding thermal-quenching activation energy $E_A$ is 8.9 meV. Similar temperature characteristics also apply to other studied GaSe SPEs: they all quench at temperatures above 100 K. Most of the GaSe PL on the pillar under relatively higher incident power density (above 1 $\mu W/\mu m^2$) exhibit improved stability with quenching above 200 K; we also notice that some localized GaSe PL is visible even at room temperature under a very high incident power density (600 $\mu W/\mu m^2$), though no antibunching is observed at those sites. While the GaSe SPEs studied in this work still require relatively low cryogenic temperature to operate, their quenching mechanisms remain elusive.

Section 11. PL spectra of the GaSe SPE at ~ 621 nm shown in Fig. 3 of the main text with different incident power densities.

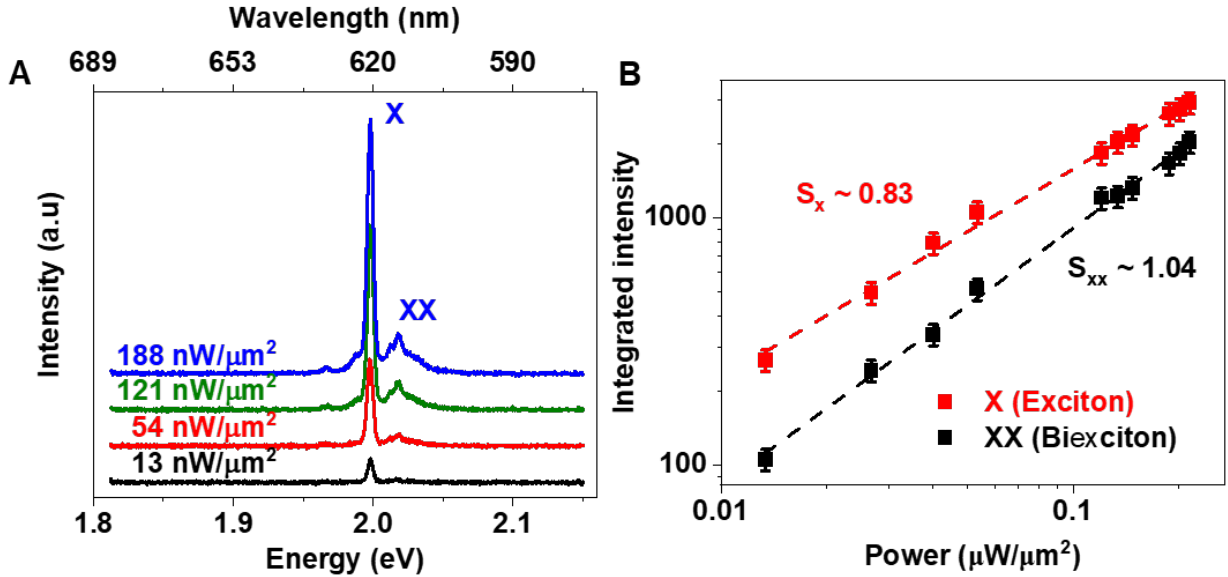

**Supplementary Fig. S15 | Power-dependent PL measurements of the SPE at ~ 621 nm are shown in Fig. 3 of the main text. (A)** The power-dependence PL spectra (measured at T = 3.5 K). **(B)** The incident power-dependent integrated intensities of exciton and biexciton of this ~ 621 nm emitter were plotted in a double-logarithmic plot.

Section 12. Quantitative analysis of biaxial tensile strain of GaSe flake on different pillars.

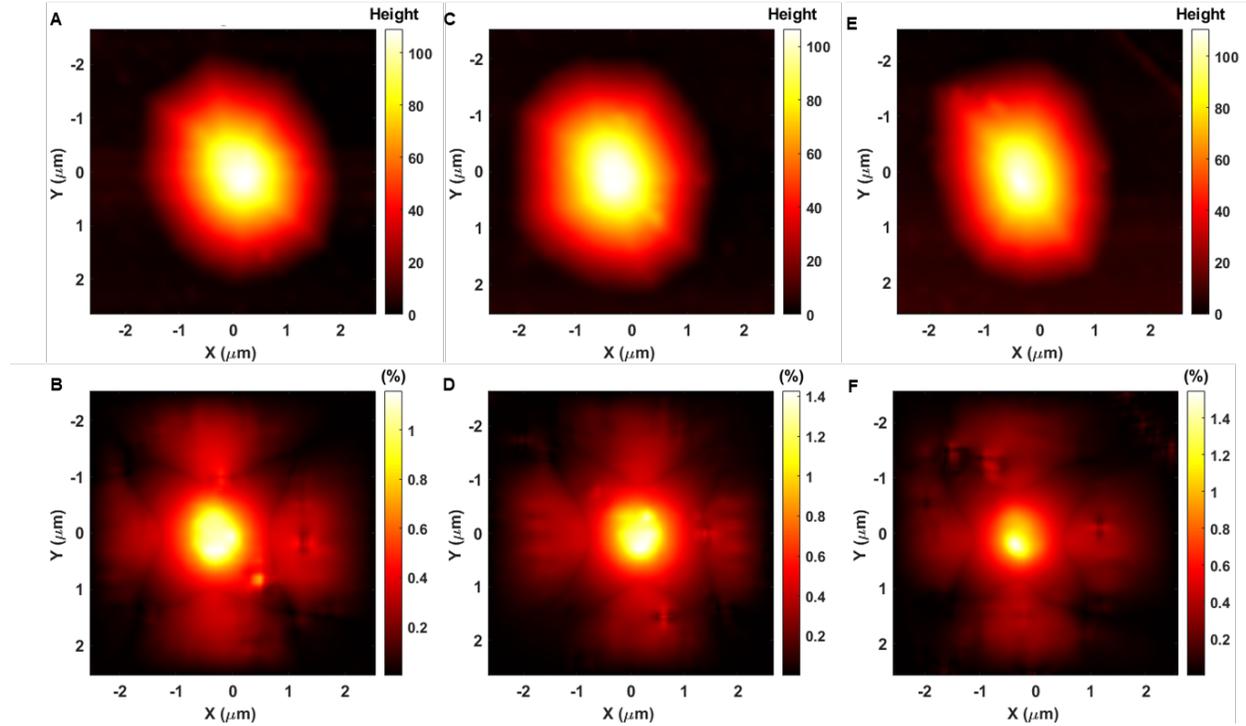

**Supplementary Fig. S16 | Simulation of strain distribution based on pillars 1 – 3 labeled in the AFM image shown in Fig. S5. (A)/(B)** represent the AFM and strain simulation results of pillar 1, respectively. **(C)/(D)** represent the AFM and strain simulation results of pillar 2, respectively. **(E)/(F)** represent the AFM and strain simulation results of pillar 3, respectively..

The classical Landau continuum model(*22*) for structural stress analysis was used to calculate the strain distribution across the pillar according to the AFM profile, which could effectively describe the nano-scale strain in nano-trenches(*23*), nano-cubes(*24*), and nano-bubbles(*25*). One can calculate the biaxial strain via the Airy stress function:

$$|\varepsilon_{total}| = \frac{E}{1-v} \left| \frac{\partial^2 H(x,y)}{\partial x^2} + \frac{\partial^2 H(x,y)}{\partial y^2} \right|$$

where E is the Young's modulus, $H(x,y)$ was the height imported from the AFM results, and $v$ = 0.23 is the Poisson ratio of $\varepsilon$-GaSe(*26, 27*). We adapted the code by Darlington. et al's(*25*) work to simulate the strain distribution across the pillars.

Section 13. Analysis of the sub-bandgap emission, strain-induced redshift of emission energy, and intensity ratio of $\frac{I_{X_S}}{I_{E_g}}$ of pillars 1 – 7 labeled in Supplementary Fig. S6.

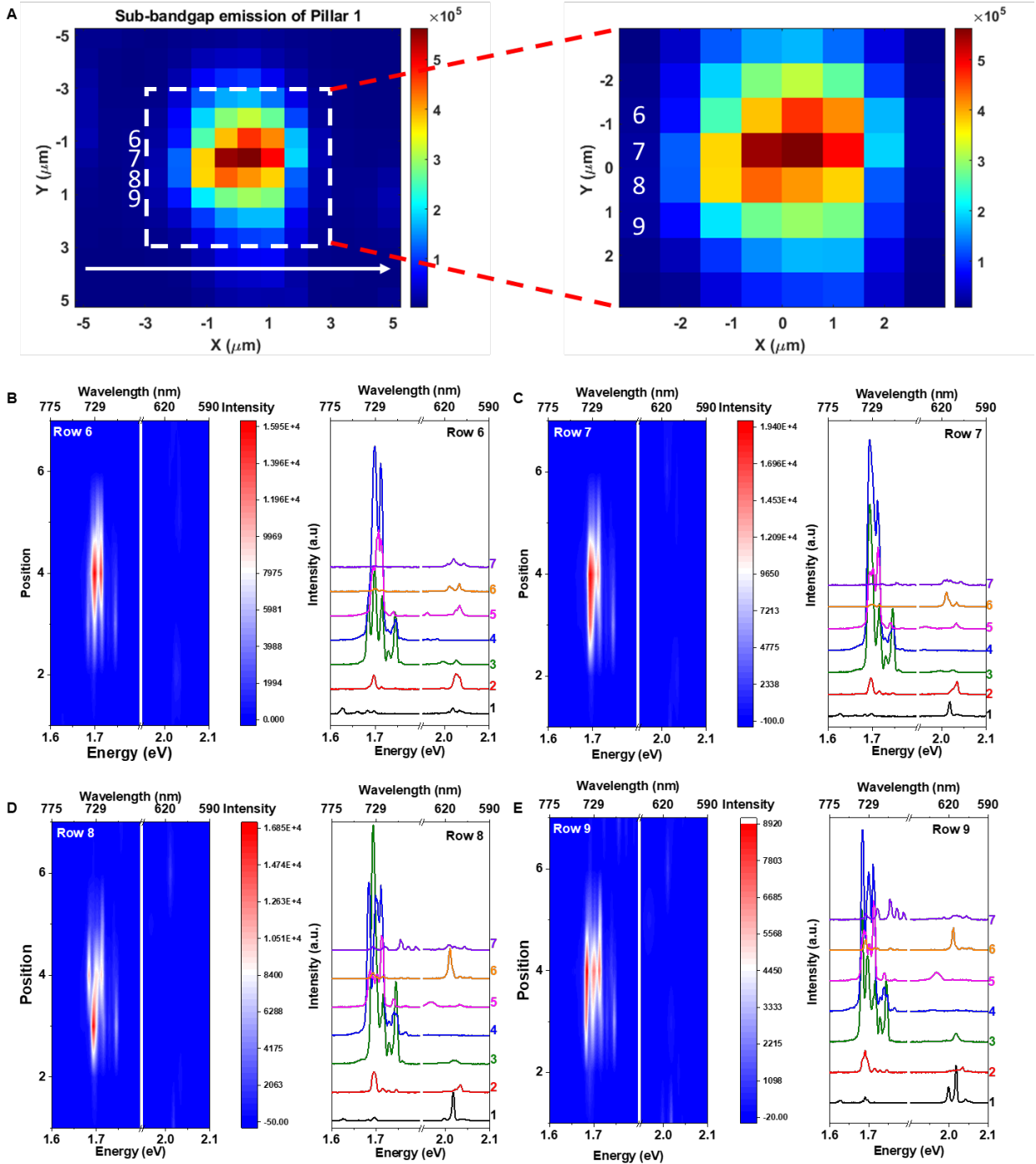

**Supplementary Fig. S17 | PL spectra (at T = 3.5 K) of different rows of pillar 1 presented in Fig. 4 of the main text. (A)** the sub-bandgap emission of pillar 1. **(B)** the PL spectra of row 6

(labeled in Fig. S17(A)). **(C)** the PL spectra of row 7 (labeled in Fig. S17(A)). **(D)** The PL spectra of row 8 (labeled in Fig. S17(A)). **(E)** The PL spectra of row 9 (labeled in Fig. S17(A)).

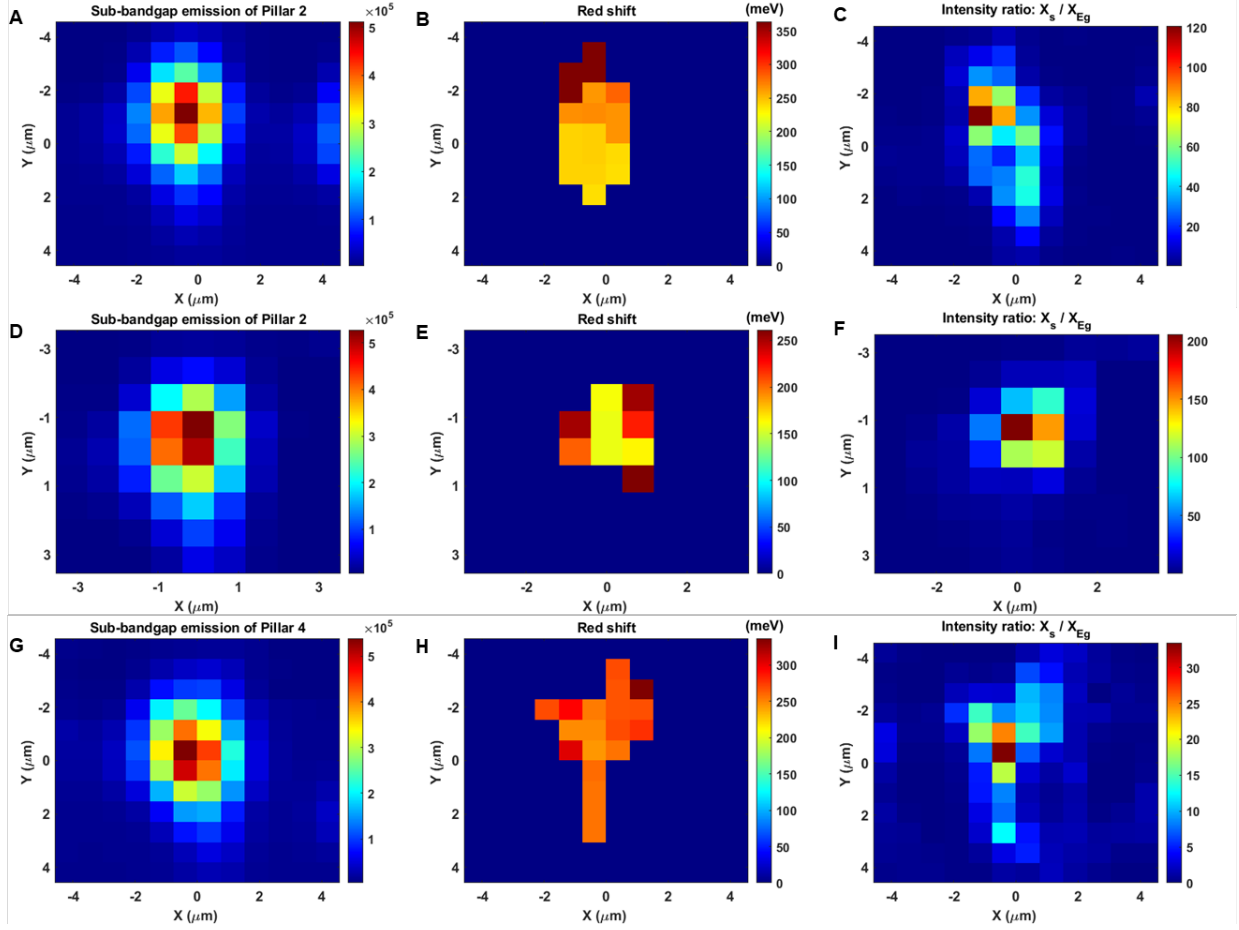

**Supplementary Fig. S18 | Analysis of the sub-bandgap emission, the strain-induced redshift of emission energy, and intensity ratio of $\frac{I_{X_S}}{I_{E_g}}$ of pillars 2 – 4. (A)-(C)** The integrated intensities (600 – 800 nm), strain-induced redshift, and position-dependent peak intensity ratios $\frac{I_{X_S}}{I_{E_g}}$ of pillar 2. **(D)-(F)** The integrated intensities (600 – 800 nm), strain-induced redshift, and peak intensity ratios $\frac{I_{X_S}}{I_{E_g}}$ of pillar 3. **(G)-(I)** The integrated intensities (600 – 800 nm), strain-induced redshift, and peak intensity ratios $\frac{I_{X_S}}{I_{E_g}}$ of pillar 4.

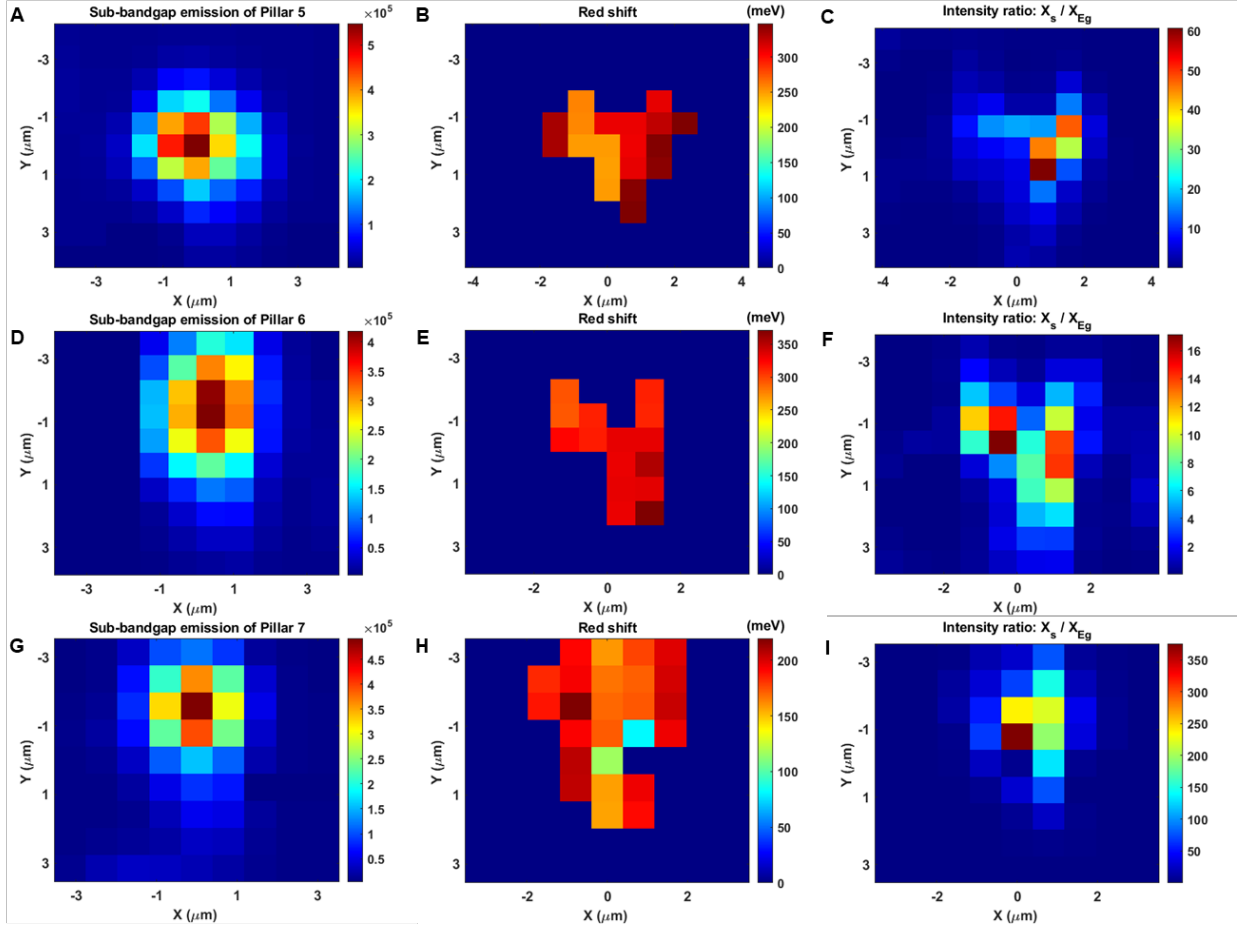

**Supplementary Fig. S19 | Analysis of the sub-bandgap emission, the strain-induced redshift of emission energy, and intensity ratio $\frac{I_{X_S}}{I_{E_g}}$ of pillars 5 – 7. (A)-(C)** The integrated intensities (600 – 800 nm), strain-induced redshift, and position-dependent peak intensity ratios $\frac{I_{X_S}}{I_{E_g}}$ of pillar 5. **(D)-(F)** The integrated intensities (600 – 800 nm), strain-induced redshift, and peak intensity ratios $\frac{I_{X_S}}{I_{E_g}}$ of pillar 6. **(G)-(I)** show the integrated intensities (600 – 800 nm), strain-induced redshift, and peak intensity ratios $\frac{I_{X_S}}{I_{E_g}}$ of pillar 7.

To probe more experimental evidence of the confinement potential, we visualize the redshift of the emission energies from the bandgap energy (~ 2.05 eV) across the pillar and correlated with strain (see Figs. S18 and S19), where the maximum redshift occurs on the pillar apex that corresponds to the maximum strain.

Section 14. Pearson's correlation coefficients between strain and the integrated intensities, redshift, and intensity ratio of $\frac{I_{X_S}}{I_{X_{E_g}}}$

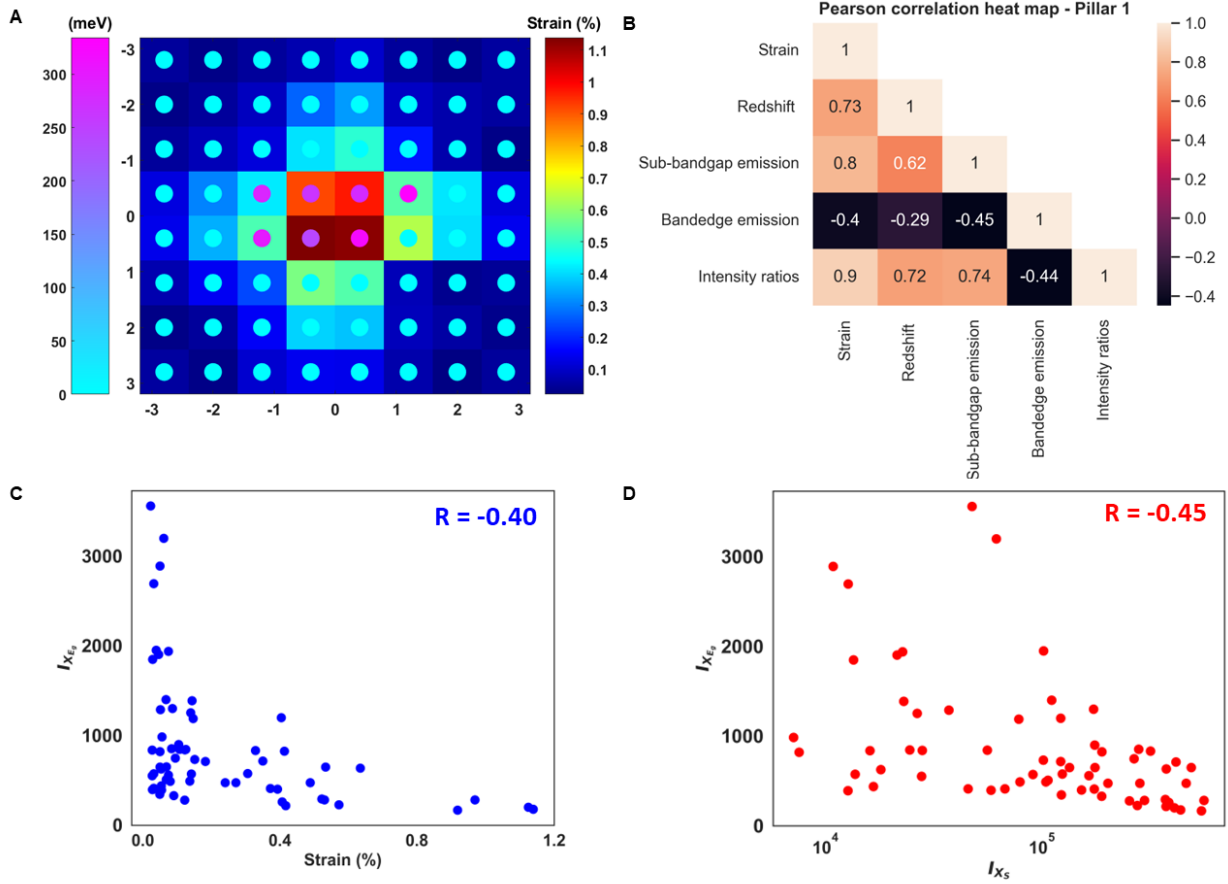

**Supplementary Fig. S20 | Correlation analysis between strain simulation results and analysis of PL spectra. (A)** Correlation map of strain gradients (squares, adapted from **Fig. 4(B)** of the main text) and the redshift (circles) between the sub-bandgap (strain-confined exciton) and band edge emission plotted on a 5.6×5.6 $\mu m$ grid. **(B)** Pearson correlation heatmap between strain and three emission properties: the integrated intensities of band edge emission $I_{X_{E_g}}$, sub-bandgap emssion $I_{X_S}$, and intensity ratio of $\frac{I_{X_S}}{I_{X_{E_g}}}$. **(C)** Pearson correlation of strain gradients and integrated intensities of band edge emission $I_{X_{E_g}}$. **(D)** Pearson correlation of integrated intensities of bandedge emission $I_{X_{E_g}}$ and integrated intensities of sub-bandgap emission $I_{X_S}$.

Note that the correlation analysis is performed on the strain gradients and the redshift of sub-bandgap emission from the band edge. Fig. S20(A) demonstrated the position-dependent redshift based on the intensity ratio of $\frac{I_{X_S}}{I_{X_{E_g}}} > 100$, implying that the larger redshift generally occurs with larger strain gradients. The correlation heatmap in Fig. S20(B) shows a strong positive correlation value of 0.73 between the strain gradients and redshift. In short, the localized strain gradients modulate the sub-bandgap emission energies.

Also, as demonstrated in Fig. 4(E) and (F) of the main text, the strain distribution of GaSe shows a strong positive correlation with the integrated intensities of sub-bandgap emission and the intensity ratio of $\frac{I_{X_S}}{I_{X_{E_g}}}$ in GaSe, which is determined by the pillar geometry. This suggests a strong correlation between the strain and the sub-bandgap emission properties that could be characterized by the Pearson coefficients. However, as shown in Fig. S20(C), the band edge emission $I_{X_{E_g}}$ shows a weak negative correlation (p = -0.40) with the strain gradients; in Fig. S20(D), the sub-bandgap emission $I_{X_S}$ also shows a weak negative (p = -0.45) correlation with the band edge emission $I_{X_{E_g}}$. This is because $I_{X_{E_g}}$ is nearly completely suppressed for larger-than-0.2% biaxial tensile strain. In short, the localized strain gradients modulate the sub-bandgap emission intensities.

Section 15. DFT calculations: Type-I band bending.

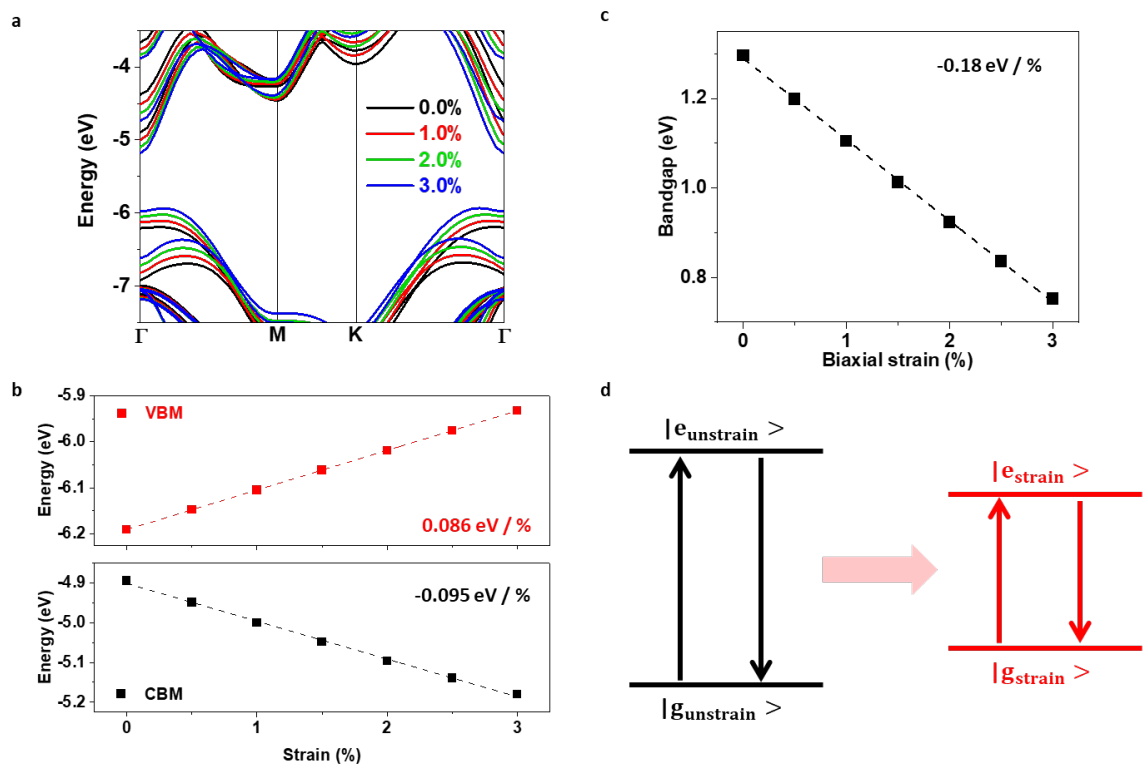

**Supplementary Fig. S21| DFT simulation results of strained GaSe. (A)** The band structures of bulk GaSe under increasing biaxial tensile strain simulated by DFT. **(B)** Change of bandgaps with 0% to 3.0% biaxial tensile strain with an increment of 0.5%. **(C)** Change of valence band maximum (VBM) and conduction band minimum (CBM) from 0% to 3.0% biaxial tensile strain with an increment of 0.5%. **(D)** Illustration of Type-I band bending.

To gain further insights on the origins of the extraordinary enhancement of sub-bandgap emission at the pillar apex, we performed band structure calculations of biaxially strained bulk GaSe for a qualitative comparison via the density functional theory (DFT) approach. Despite the common underestimation of the bandgap with DFT, it still reflects the correct physical trend. As shown in Fig. S21(A) and (B), the bandgap constantly reduces with increasing biaxial tensile strain at -0.18 eV/%. Fig. S21(C) demonstrates the changes in the valence band maximum (VBM) and conduction band minimum (CBM): the VBM and CBM show different bending directions with rates of 0.086 eV/% and -0.095 eV/%. Such opposite band bending associated with localized nanostructure could be well described as "Type-I funneling"(*28*), where the energy level of excited electrons descends towards the pillar apex with a larger strain while that of holes ascends. As a result, the charge carriers are concentrated at the apex that facilitates exciton emission. Hence, the brightening of sub-bandgap emission at the pillar apex could be attributed to the strain-induced redshift of the bandgap (confinement potential) and "Type-I funneling." The energy state diagram in Fig. S21(D) suggests the bandgap narrowing caused by the increasing strain that matches the redshift of emission energy observed experimentally.